\documentclass[12pt]{article}
\pdfoutput=1  
\usepackage[margin=0.95 in]{geometry}
\usepackage{amsmath}
\usepackage{amssymb,amsfonts}
\usepackage[all]{xy}
\usepackage{graphicx}
\usepackage[utf8x]{inputenc}
\usepackage{amsmath}
\usepackage{amssymb}
\usepackage{float}
\usepackage{array}
\usepackage{tikz}
\usepackage{mathtools}
\usepackage{mathrsfs} 
\usepackage[hidelinks]{hyperref}
\usepackage{cite}
\usepackage{tikz-cd}
\usepackage{graphicx}
\usepackage{bm}
\usepackage{rotating}
\usepackage{array}
\usepackage{xcolor}
\usepackage{amsmath,amssymb}
\usepackage{mathrsfs}
\usepackage{graphicx}
\usepackage{color}
\usepackage{subfigure}
\usepackage{fancyhdr}
\usepackage{multirow}
\usepackage{float}
\usepackage{epsfig}
\usepackage{amsfonts}
\usepackage{bm}

\newcommand{\ba}{\begin{eqnarray}}
\newcommand{\ea}{\end{eqnarray}}

\numberwithin{equation}{section}
\numberwithin{equation}{section}
\numberwithin{table}{section}

\begin{document}

\vspace*{3cm}{}
	
	\noindent
	{\LARGE \bf  Inflation from $f(R)$ theories in gravity's rainbow}
%	\vskip.1cm
%	\noindent
%	{\LARGE \bf  Lie Algebra Cohomology}
	\vskip .4cm
	\noindent
	\linethickness{.04cm}
	\line(10,0){467}
	\vskip 1cm
	\noindent
	{\large \bf  Areef Waeming$^{1,\dagger}$, Phongpichit Channuie$^{1,2,3,4,\ddagger}$}
\vskip 0.5cm
{\em 
\noindent
$^1$School of Science, Walailak University, Nakhon Si Thammarat, 80160, Thailand \\
$^2$College of Graduate Studies, Walailak University, Nakhon Si Thammarat, 80160, Thailand\\
$^3$Research Group in Applied, Computational and Theoretical Science (ACTS), \\Walailak University, Nakhon Si Thammarat, 80160, Thailand\\
$^4$Thailand Center of Excellence in Physics, Ministry of Higher Education, Science, \\Research and Innovation, Bangkok 10400, Thailand\\
{\rm Email:}\,$^{\dagger}${\rm reef.waeming@gmail.com},\,$^{\ddagger}${\rm channuie@gmail.com}}

\vskip 1cm

\noindent {\sc Abstract: } In this work, we study the $f(R)$ models of inflation in the context of gravity's rainbow theory. We choose three types of $f(R)$ models: $f(R)=R+\alpha (R/M)^{n},\,f(R)=R+\alpha R^{2}+\beta R^{2}\log(R/M^{2})$ and the Einstein-Hu-Sawicki model with $n,\,\alpha,\,\beta$ being arbitrary real constants. Here $R$ and $M$ are the Ricci scalar and mass scale, respectively. For all models, the rainbow function is written in the power-law form of the Hubble parameter. We present a detailed derivation of the spectral index of curvature perturbation and the tensor-to-scalar ratio and compare the predictions of our results with latest Planck 2018 data. With the sizeable number of e-foldings and proper choices of parameters, we discover that the predictions of all $f(R)$ models present in this work are in excellent agreement with the Planck analysis.
	
\vskip 1cm
%\pagebreak
	
\newpage
\setcounter{tocdepth}{2}
\tableofcontents

%%%%%%%%%%%%%%%%%%%%%%%%%%%%%%%%%%%%%%%%

\section{Introduction}
As an effective theory of gravity, Einstein's general theory of gravity is valid in the low energy regime (IR), while at very high energy regime (UV) the Einstein theory could in principle be improved. It is expected that the usual dispersion relation will get modified at the energy scale of the order of Planck length in various theories of quantum gravity. Deformations of dispersion relations are timely since the Cherenkov array \cite{Cherenkov} will focus on this type of deformations as well. In Ref.\cite{Magueijo:2002xx}, Magueijo and
Smolin introduced a modification of the dispersion relation by replacing the standard one, i.e. $\epsilon^{2}-p^{2}=m^{2}$, replaced with a new form $\epsilon^{2}{\tilde f}^{2}(\epsilon) - p^{2}{\tilde g}^{2}(\epsilon)=m^{2}$ where ${\tilde f}(\epsilon)$ and ${\tilde g}(\epsilon)$ are so-called rainbow functions. They are required to satisfy standard properties at a low-energy IR limit when $\epsilon/M \rightarrow 0$ that ${\tilde f}(\epsilon/M) \rightarrow 1$ and ${\tilde g}(\epsilon/M)\rightarrow 1$ where $M$ is the energy scale that quantum effects of gravity become relevant.   

It is expected that the usual dispersion relation in the UV limit has to be in principle reformed and captures a modification of the geometry at that limit. One way to think about this is to assume that the geometry of the space-time in gravity's rainbow depends on energy of the test particles. Therefore, each test particle carrying different energy will feel a different geometry of space-time. This behavior promotes a family of metrics, namely a rainbow metric, characterized by $\epsilon$ to describe the background of the space-time instead of a single metric. In gravity's rainbow, the modified metric can be expressed as
\ba
g(\epsilon) = \eta^{\mu\nu}{\tilde e}_{\mu}(\epsilon)\otimes{\tilde e}_{\nu}(\epsilon)\,,
\label{act}
\ea
where the energy dependence of the frame field ${\tilde e}_{\mu}(\epsilon)$ can be written in terms of the energy independence
frame field $e_{\mu}$ as ${\tilde e}_{0}(\epsilon)= e_{0}/{\tilde f}(\epsilon)$ and ${\tilde e}_{i}(\epsilon)= e_{i}/{\tilde g}(\epsilon)$ where $i=1,2,3$. In the cosmological viewpoint, the conventional FLRW spacetime metric for homogeneous and isotropic universe is replaced by a rainbow metric of the form
\begin{eqnarray}
{ds}^2(\epsilon) =-\frac{dt^2}{\tilde{f}^2(\epsilon)}+\frac{a^2(t)}{\tilde{g}^2(\epsilon)} \delta_{ij}{dx}^i{dx}^j\,,\label{modFR}
\end{eqnarray}
where $a(t)$ is a scale factor. In Ref.\cite{Ling:2006az}, the author studied the semi-classical effect
of radiation particles on the background metric in the
framework of rainbow gravity, and obtained the modified FLRW equations where the energy of particles varies
with the cosmological time. In the present work, for very early universe, we consider the rainbow functions ${\tilde f}(\epsilon)$ dependent on time implicitly through
the energy of particles. 

In recent years, a framework of gravity's rainbow has attracted a lot of attentions and became the subject of much interest
in the literature. Among many relevant publications, various physical properties of the black holes have been investigated, see e.g. \cite{Feng:2016zsj,Hendi:2016hbe,Feng:2017gms,Hendi:2018sbe,Panahiyan:2018fpb,Dehghani:2018qvn,Upadhyay:2018vfu,Dehghani:2018svw,Hendi:2017pld,Hendi:2016njy,Hendi:2015cra,Hendi:2015hja,Ali:2015iba,Ali:2014zea,Ali:2014qra,Ali:2014yea,Ali:2014cpa,EslamPanah:2018ums}. Moreover, the effects of the rainbow functions have also been discussed in other cosmological scenarios, see for instance \cite{Momeni:2017cvl,Deng:2017umx,Khodadi:2016aop,Khodadi:2016bcx,Rudra:2016alu,Ashour:2016cay,Garattini:2012ec,Garattini:2014rwa}. It was found that the gravity's rainbow was investigated in Gauss-Bonnet gravity \cite{Hendi:2016tiy}, massive gravity \cite{Hendi:2017vgo,Heydarzade:2017rpb} and $f(R)$ gravity \cite{Hendi:2016oxk}. More specifically, the gravity's rainbow has also been used for analyzing the effects of rainbow functions
on the Starobinsky model of $f(R)$ gravity \cite{Chatrabhuti:2015mws}. More recently, the deformed Starobinsky model \cite{Codello:2014sua} has been also studied in the context of gravity's rainbow \cite{Channuie:2019kus}.

This paper is organized as follows: In section \ref{sec1}, we formulate $f(R)$ theory \cite{Sotiriou:2008rp,DeFelice:2010aj} in the framework of gravity's rainbow. We study the $f(R)$ models of inflation in the context of gravity's rainbow theory by considering three types of $f(R)$ models: $f(R)=R+\alpha (R/M)^{n},\,f(R)=R+\alpha R^{2}+\beta R^{2}\log(R/M^{2})$ and the Einstein-Hu-Sawicki model. We take a short review of a cosmological linear perturbation in the context of the gravity's rainbow generated during inflation. Here we compute the spectral index
of scalar perturbation and the tensor-to-scalar ratio of
the model in section \ref{sec3}. In section \ref{sec4}, we compare the predicted results with Planck 2018 data. We conclude our findings in the last section. In this work, we use the metric signature $(-,+,+,+)$. The Greek indices $\mu, \nu, \rho, ...$ run from $0$ to $3$, whereas the Latin indices $i, j, k, ...$ run from $1$ to $3$ (spatial components).

\section{General Setup: $f(R)$ gravity's rainbow}
\label{sec1}
As is well known, Einstein's theory of gravitation remains a pillar of modern description of gravity as a fundamental property of space-time. However, the theory itself is plagued by major unsolved problems nowadays, e.g. dark matter, dark energy and even cosmic inflation. Hence, the modification to general relativity are expected to be plausible in the very early universe where possible corrections to Einstein's theory may in principle appear at high curvature. One of the simplest classes of such modifications is to replace the Einstein-Hilbert term in the action with a generic function of the Ricci scalar. This class of theories is known as the $f(R)$ theories. Note that there were much earlier and pioneer works on $f(R)$ and other gravity theories, see \cite{Nojiri:2010wj,Nojiri:2017ncd}. Here we initiate our setup with the traditionally 4-dimensional action in $f(R)$ gravity including the matter fields as \cite{Sotiriou:2008rp,DeFelice:2010aj}. 
\ba
S = \frac{1}{2\kappa^2}\int d^4x \sqrt{-g}f(R) + \int d^4x \sqrt{-g}{\cal L}_{M}(g_{\mu\nu},\Psi_{M})\,,
\label{act}
\ea
where we have defined $\kappa^{2}=8\pi G=8\pi/m^{2}_{\rm P}$, $g$ is the determinant of the metric $g_{\mu\nu}$, and the matter field Lagrangian ${\cal L}_{M}$ depends on $g_{\mu\nu}$ and matter fields $\Psi_{M}$. The field equation can be directly derived by performing variation of the action (\ref{act}) with respect to $g_{\mu\nu}$ to obtain \cite{Sotiriou:2008rp,DeFelice:2010aj}
\ba
F(R)R_{\mu\nu}(g) - \frac{1}{2}f(R)g_{\mu\nu}-\nabla_{\mu}\nabla_{\nu}F(R)+g_{\mu\nu}\Box F(R) = \kappa^{2}T^{(M)}_{\mu\nu}\,,
\label{eom}
\ea
where $F(R)=\partial f(R)/\partial R$ and the operator $\Box$ is defined by $\Box\equiv (1/\sqrt{-g})\partial_{\mu}(\sqrt{-g}g^{\mu\nu}\partial_{\nu})$. Basically, the energy-momentum tensor of the matter fields is given by a definition $T^{(M)}_{\mu\nu}=(-2/\sqrt{-g})\delta(\sqrt{-g}{\cal L}_{M})/\delta g^{\mu\nu}$. Here it satisfies the continuity equation such that $\nabla^{\mu}T^{(M)}_{\mu\nu}=0$. As a part the standard approach, it is worth noting that the energy-momentum tensor of matter is given in the perfect fluid form as $T^{(M)}_{\mu\nu}={\rm diag}(-\rho,\,P,\,P,\,P)$ with $\rho$ and $P$ being the energy density and the pressure, respectively. Now we are going to derive cosmological solutions to field equations (\ref{eom}).  Inserting the modified FLRW metric (\ref{modFR}) in the field equations (\ref{eom}) and assuming that the
stress-energy tensor is written in terms of the perfect fluid form yield
\begin{eqnarray}
3 \left(FH^2+H\dot{F} \right) -6FH\frac{  \dot{\tilde{g}}}{\tilde{g}} +3 F\frac{ \dot{\tilde{g}}^2}{\tilde{g}^2}+\dot{F}\frac{ \dot{\tilde{f}}}{\tilde{f}}-3 \dot{F}\frac{ \dot{\tilde{g}}}{\tilde{g}}=\frac{F R-f(R)}{2 \tilde{f}^2}+\frac{\kappa ^2 \rho }{\tilde{f}^2}\,,\label{Hdd}
\end{eqnarray}
and
\begin{eqnarray}
&&3 F H^2-3 \dot{F} H+3 F \dot{H}+3 F H\frac{ \dot{\tilde{f}}}{\tilde{f}}-\dot{F}\frac{ \dot{\tilde{f}}}{\tilde{f}}-4 F\frac{ \dot{\tilde{g}}^2}{\tilde{g}^4}+6 F H\frac{ \dot{\tilde{g}}}{\tilde{g}^3}-3 \dot{F}\frac{ \dot{\tilde{g}}}{\tilde{g}^3}+F\frac{ \ddot{\tilde{g}}}{\tilde{g}^3}+F\frac{\dot{\tilde{f}} }{\tilde{f}}\frac{\dot{\tilde{g}}}{\tilde{g}^3}-3 F H^2\frac{1}{\tilde{g}^2}\nonumber\\&&+2 \dot{F} H\frac{1}{\tilde{g}^2}+\frac{\ddot{F}}{\tilde{g}^2}-F \dot{H}\frac{1}{\tilde{g}^2}+6 F\frac{ \dot{\tilde{g}}^2}{\tilde{g}^2}-F H\frac{ \dot{\tilde{f}}}{\tilde{f} }\frac{1}{\tilde{g}^2}+\dot{F}\frac{ \dot{\tilde{f}}}{\tilde{f}}\frac{1}{\tilde{g}^2}-6 F H\frac{ \dot{\tilde{g}}}{\tilde{g}}+3 \dot{F}\frac{ \dot{\tilde{g}}}{\tilde{g}}-3 F\frac{ \ddot{\tilde{g}}}{\tilde{g}}-3 F\frac{\dot{\tilde{f}} }{\tilde{f} }\frac{\dot{\tilde{g}}}{\tilde{g}} \nonumber\\&&-\frac{f(R) \left(\tilde{g}-1\right) \left(\tilde{g}+1\right)}{2 \tilde{f}^2 \tilde{g}^2}=-\frac{\kappa ^2 \left(\rho  \tilde{g}^2+P\right)}{\tilde{f}^2 \tilde{g}^2}\,,\label{ijcom}
\end{eqnarray}
where we have defined a first and second derivative with respect to time with ${\dot a}$ and ${\ddot a}$, respectively. For simplicity, in our analysis below we chose $\tilde{g}=1$ and only considered the spatially flat universe. Detailed calculations of the Ricci scalar are given in Appendix \ref{apa}.

\subsection{Model I: $f(R)=R+\alpha (R/M)^{n}$}\label{21}
Modification of the Einstein–Hilbert (EH) action through the
Ricci scalar can describe early universe, e.g. cosic inflation. In this first model, we start by considering a simple case of the $f(R)$ gravity model where the Ricci scalar replaced by a new function as
\begin{eqnarray}
f(R)=R+\alpha \Big(\frac{R}{M}\Big)^{n}, \label{star}
\end{eqnarray}
where $\alpha>0$ and $n$ are arbitrary constants assuming that $n\geq 2$. Notice that when setting $\alpha=1/6$ and $n=2$ we obtain $f(R)$ gravity of Starobinsky model \cite{Starobinsky:1980te}. From Eq.(\ref{Hdd}), we find for this model
\begin{eqnarray}
&&3 H^2+\frac{\alpha  6^{n-1}}{\tilde{f}^3 \big(\big(2 H^2+\dot{H}\big) \tilde{f}+H \dot{\tilde{f}}\big)^2} \Big(\frac{\tilde{f} \big(\big(2 H^2+\dot{H}\big) \tilde{f}+H \dot{\tilde{f}}\big)}{M}\Big)^n \Big(H (n-1) n \dot{\tilde{f}}^3\nonumber\\&&+(n-1) \dot{\tilde{f}} \tilde{f} \big(\dot{\tilde{f}} \big(H^2 (7 n-3)+3 \dot{H} n\big)+H n \ddot{\tilde{f}}\big)+\big((n-1) n \dot{\tilde{f}} \ddot{H}+H\big(\big(3 H^2 (n (4 n-7)+4)\nonumber\\&&+\dot{H} (n-1) (13 n-6)\big)\dot{\tilde{f}}+3 H (n-1) n \ddot{\tilde{f}}\big)\big)\tilde{f}^2+3\big(-2 H^4 (n-2)+H^2 \dot{H} (n (4 n-7)+4)\nonumber\\&&+H (n-1) n \ddot{H}+\dot{H}^2 (-(n-1))\big)\tilde{f}^3\Big)=0,\label{1sft}
\end{eqnarray}
and
\begin{eqnarray}
&&\frac{1}{3 R^3 \tilde{f}^2}\Bigg(\alpha  (n-1) n R \tilde{f} \Big(\frac{R}{M}\Big)^n \big(\dot{R} \big(9 H \tilde{f}+4 \dot{\tilde{f}}\big)+3 \tilde{f} \ddot{R}\big)+3 \alpha  (n-2) (n-1) n \dot{R}^2 \tilde{f}^2 \Big(\frac{R}{M}\Big)^n\nonumber\\&&-\alpha  (n-2) R^3 \Big(\frac{R}{M}\Big)^n+R^4\Bigg)=0. \label{2sft}
\end{eqnarray}
Here we are only interested in an inflationary solution. Therefore we invoke the slow-roll approximations. Hence the terms containing $\ddot{H}$ and higher power in $\dot{H}$ can be neglected in this particular regime. It is rather straightforward to show that the Eq.(\ref{1sft}) is reduced to
\begin{eqnarray}
&\dot{H}&\simeq\frac{H^2 2^{1-2 n} 3^{-n}}{\alpha  (\lambda +1) \left(4 n^2-7 n+4\right)} \left(\frac{H^2 \left(\frac{H}{M}\right)^{2 \lambda }}{M}\right)^{-n} \nonumber\\&&\Bigg(-12 H^2 \left(\frac{H}{M}\right)^{2 \lambda }+\alpha  \left(-2^{2 n+1}\right) 3^n \left(\frac{H^2 \left(\frac{H}{M}\right)^{2 \lambda }}{M}\right)^n+\alpha  2^{2 n} 3^n n \Big(\frac{H^2 \left(\frac{H}{M}\right)^{2 \lambda }}{M}\Big)^n\Bigg).\label{HdHumod1}
\end{eqnarray}
Note that when setting $n=2$ and $\alpha=1/6$ the result converts to that of Ref.\cite{Chatrabhuti:2015mws}. During inflation we can assume $H\simeq {\rm constant.}$, and then in this situation we obtain from Eq.(\ref{HdHumod1})
\ba
H &\simeq& H_i +\frac{H^2_i 2^{1-2 n} 3^{-n}}{\alpha  (\lambda +1) \left(4 n^2-7 n+4\right)} \left(\frac{H^2_i \left(\frac{H_i}{M}\right)^{2 \lambda }}{M}\right)^{-n}\Bigg(-12 H^2_i \left(\frac{H_i}{M}\right)^{2 \lambda }\nonumber\\&&\quad\quad+\alpha  \left(-2^{2 n+1}\right) 3^n \left(\frac{H^2_i \left(\frac{H_i}{M}\right)^{2 \lambda }}{M}\right)^n+\alpha  2^{2 n} 3^n n \Big(\frac{H^2 \left(\frac{H_i}{M}\right)^{2 \lambda }}{M}\Big)^n\Bigg)(t-t_i)\,,\label{Htkmod1}
\ea
and 
\ba
a &\simeq& a_i \exp \Bigg\{H_i(t-t_i) +\frac{H^2_i 2^{1-2 n} 3^{-n}}{\alpha  (\lambda +1) \left(4 n^2-7 n+4\right)} \left(\frac{H^2_i \left(\frac{H_i}{M}\right)^{2 \lambda }}{M}\right)^{-n}\Bigg(-12 H^2_i \left(\frac{H_i}{M}\right)^{2 \lambda }\nonumber\\&&\quad\quad\quad+\alpha  \left(-2^{2 n+1}\right) 3^n \left(\frac{H^2_i \left(\frac{H_i}{M}\right)^{2 \lambda }}{M}\right)^n+\alpha  2^{2 n} 3^n n \Big(\frac{H^2 \left(\frac{H_i}{M}\right)^{2 \lambda }}{M}\Big)^n\Bigg)\frac{(t-t_i)^{2}}{2}\Bigg\},
\ea
where $H_i$ and $a_i$ are respectively the Hubble parameter and the scale factor at the onset of inflation ($t=t_i$). The slow-roll parameter $\epsilon_1$ is defined by $\epsilon_1 \equiv -\dot{H}/H^2$ which in this case can be estimated as
\begin{eqnarray}
\epsilon_{1}\equiv -\frac{\dot{H}}{H^{2}}&\simeq& -\frac{2^{1-2 n} 3^{-n}}{\alpha  (\lambda +1) \left(4 n^2-7 n+4\right)} \left(\frac{H^2 \left(\frac{H}{M}\right)^{2 \lambda }}{M}\right)^{-n}\Bigg(-12 H^2 \left(\frac{H}{M}\right)^{2 \lambda }\nonumber\\&&\quad\quad+\alpha  \left(-2^{2 n+1}\right) 3^n \left(\frac{H^2 \left(\frac{H}{M}\right)^{2 \lambda }}{M}\right)^n+\alpha  2^{2 n} 3^n n \Big(\frac{H^2 \left(\frac{H}{M}\right)^{2 \lambda }}{M}\Big)^n\Bigg).\label{epumod1}
\end{eqnarray}
We can check that $\epsilon_{1}$ is less than unity during inflation ($H^2 \gg M^2$) and we find when setting $n=2,\,\alpha=1/6$ that the bove expression reduces to $\epsilon_1 \simeq\frac{H^{-2 (\lambda +1)} M^{2 \lambda +2}}{6 (\lambda +1)}$. One can simply determine the time when inflation ends ($t=t_f$) by solving $\epsilon(t_f) \simeq 1$ to obtain
\ba
t_f &\simeq& t_i -\Bigg[\frac{H_i 2^{1-2 n} 3^{-n}}{\alpha  (\lambda +1) \left(4 n^2-7 n+4\right)}\Bigg]^{-1} \left(\frac{H^2_i \left(\frac{H_i}{M}\right)^{2 \lambda }}{M}\right)^{n}\Bigg(-12 H^2_i \left(\frac{H_i}{M}\right)^{2 \lambda }\nonumber\\&&\quad\quad\quad+\alpha  \left(-2^{2 n+1}\right) 3^n \left(\frac{H^2_i \left(\frac{H_i}{M}\right)^{2 \lambda }}{M}\right)^n+\alpha  2^{2 n} 3^n n \Big(\frac{H^2 \left(\frac{H_i}{M}\right)^{2 \lambda }}{M}\Big)^n\Bigg)^{-1}\, . \label{tfmod1}
\ea
The number of e-foldings from $t_i$ to $t_f$ is then given by
\ba
N &\equiv& \int^{t_f}_{t_i} Hdt \simeq H_i(t-t_i)\nonumber\\&& +\frac{H^2_i 2^{1-2 n} 3^{-n}}{\alpha  (\lambda +1) \left(4 n^2-7 n+4\right)} \left(\frac{H^2_i \left(\frac{H_i}{M}\right)^{2 \lambda }}{M}\right)^{-n}\Bigg(-12 H^2_i \left(\frac{H_i}{M}\right)^{2 \lambda }\nonumber\\&&+\alpha  \left(-2^{2 n+1}\right) 3^n \left(\frac{H^2_i \left(\frac{H_i}{M}\right)^{2 \lambda }}{M}\right)^n+\alpha  2^{2 n} 3^n n \Big(\frac{H^2 \left(\frac{H_i}{M}\right)^{2 \lambda }}{M}\Big)^n\Bigg)\frac{(t-t_i)^{2}}{2}\nonumber\\&\simeq& \frac{1}{2\epsilon_{1}(t_{i})}\, .
\ea
Note that when $c_{1}=-1/6,\,c_{2}=0$ and $\lambda=0$ the result is the same as that of the Starobinsky model.

\subsection{Model II: $f(R)=R+\alpha R^{2} + \beta R^{2}\log(R/M^{2})$}
In this second model, we consider the particular case of the $f(R)$ gravity model where the Ricci scalar replaced by a new function as
\begin{eqnarray}
f(R)=R+\alpha R^{2} + \beta R^{2}\log\Big(\frac{R}{M^{2}}\Big), \label{loa}
\end{eqnarray}
where $M$ is a mass scale and also $\alpha,\,\beta >0$. Note here that when setting $\beta=0$ and $\alpha=1/(2M^{2})$, this model is reduced to the Starobinsky model. In a very similar $f(R)$ form, a logarithmic-corrected $R^{2}$ model was
considered in Refs.\cite{Nojiri:2003ni,Nojiri:2010wj,Elizalde:2018now}. It was also found in Ref.\cite{Shamir:2019pmu} that the same corrected form has been used in studying compact stars. From the equation (\ref{loa}), we obtain
\begin{eqnarray}
f'(R)\equiv \frac{\partial f(R)}{\partial R}&=& 1+{\tilde \gamma} R + 2\beta R\log\Big(\frac{R}{M^{2}}\Big)\,,\\
f''(R)\equiv \frac{\partial^{2} f(R)}{\partial R^{2}}&=&{\tilde\lambda} + 2\beta R\log\Big(\frac{R}{M^{2}}\Big)\,,
\end{eqnarray}
where ${\tilde \gamma}=2\alpha + \beta$ and ${\tilde\lambda}=2\alpha + 3\beta$. As mentioned in Ref.\cite{Sadeghi:2016jfv}, a natural logarithmic correction is necessary to have cosmological
parameters in agreement with the recent Plank 2015 results. The function $f(R)$ obeys the quantum stability condition $f''(R)>0$ for $\alpha >0$ and $\beta >0$. This ensures the stability of the solution at high curvature. Additionally, the condition of classical stability leads to
\begin{eqnarray}
f'(R) = 1+{\tilde \gamma} R + 2\beta R\log\Big(\frac{R}{M^{2}}\Big)>0\,.
\end{eqnarray}
From Eq.(\ref{Hdd}), we find for this model
\begin{eqnarray}
&&3\Bigg(6 \dot{H} \dot{\tilde{f}}^2 \big(2 \beta  \log \Big(\frac{6 \tilde{f} \big(\big(2 H^2+\dot{H}\big) \tilde{f}+H \dot{\tilde{f}}\big)}{M^2}\Big)+2 \alpha +3 \beta \big)\nonumber\\&&+18 H^3 \dot{\tilde{f}} \tilde{f} \big(2 \beta  \log \Big(\frac{6 \tilde{f} \big(\big(2 H^2+\dot{H}\big) \tilde{f}+H \dot{\tilde{f}}\big)}{M^2}\Big)+2 \alpha +3 \beta \big)\nonumber\\&&+2 \dot{\tilde{f}} \tilde{f} \ddot{H} \big(2 \beta  \log \Big(\frac{6 \tilde{f} \big(\big(2 H^2+\dot{H}\big) \tilde{f}+H \dot{\tilde{f}}\big)}{M^2}\Big)+2 \alpha +3 \beta \big)\nonumber\\&&-12 \beta  H^4 \tilde{f}^2-6 \dot{H}^2 \tilde{f}^2 \big(\beta  \log \Big(\frac{6 \tilde{f} \big(\big(2 H^2+\dot{H}\big) \tilde{f}+H \dot{\tilde{f}}\big)}{M^2}\Big)+\alpha +\beta \big)\nonumber\\&&+H^2\Big(\dot{\tilde{f}}^2 \big(22 \beta  \log \Big(\frac{6 \tilde{f} \big(\big(2 H^2+\dot{H}\big) \tilde{f}+H \dot{\tilde{f}}\big)}{M^2}\Big)+22 \alpha +36 \beta \big)+1\big)\nonumber\\&&+6 \tilde{f} \big(3 \dot{H} \tilde{f}+\ddot{\tilde{f}}\big) \big(2 \beta  \log \Big(\frac{6 \tilde{f} \big(\big(2 H^2+\dot{H}\big) \tilde{f}+H \dot{\tilde{f}}\big)}{M^2}\Big)+2 \alpha +3 \beta \big)\Big)\nonumber\\&&+\frac{1}{\tilde{f}}2 H\Big(\dot{\tilde{f}}^3 \big(2 \beta  \log \Big(\frac{6 \tilde{f} \big(\big(2 H^2+\dot{H}\big) \tilde{f}+H \dot{\tilde{f}}\big)}{M^2}\Big)+2 \alpha +3 \beta \big)\nonumber\\&&+3 \tilde{f}^3 \ddot{H} \big(2 \beta  \log \Big(\frac{6 \tilde{f} \big(\big(2 H^2+\dot{H}\big) \tilde{f}+H \dot{\tilde{f}}\big)}{M^2}\Big)+2 \alpha +3 \beta \big)\nonumber\\&&+\dot{\tilde{f}}\tilde{f}\big(\ddot{\tilde{f}} \big(2 \beta  \log \Big(\frac{6 \tilde{f} \big(\big(2 H^2+\dot{H}\big) \tilde{f}+H\dot{\tilde{f}}\big)}{M^2}\Big)+2 \alpha +3 \beta \big)\nonumber\\&&+\dot{H} \tilde{f} +\big(20 \beta  \log \Big(\frac{6 \tilde{f} \big(\big(2 H^2+\dot{H}\big) \tilde{f}+H \dot{\tilde{f}}\big)}{M^2}\Big)+20 \alpha +33 \beta \big)\big)\Big)\Bigg)=0\,,\label{2sft}
\end{eqnarray}
and
\begin{eqnarray}
&&\frac{1}{3 R \tilde{f}^2}\Big(R \tilde{f} \big(\dot{R} \big(9 H \tilde{f}+4 \dot{\tilde{f}}\big)+3 \tilde{f} \ddot{R}\big) \big(2 \alpha +3 \beta +2 \beta  \log \Big(\frac{R}{M^2}\Big)\big)\nonumber\\&&+6 \beta  \dot{R}^2 \tilde{f}^2-\beta  R^3+R^2\Big)=0\,. \label{22sft}
\end{eqnarray}
Here we are only interested in an inflationary solution. Therefore we invoke the slow-roll approximations. Hence the terms containing $\ddot{H}$ and higher power in $\dot{H}$ can be neglected in this particular regime. It is rather straightforward to show that the Eq.(\ref{2sft}) is reduced to
\begin{eqnarray}
&\dot{H}&\simeq \frac{\left(\frac{H}{M}\right)^{-2 \lambda } \left(12 \beta  H^2 \left(\frac{H}{M}\right)^{2 \lambda }-1\right)}{18 (\lambda +1) \left(2 \alpha +3 \beta +2 \beta  \log \left(\frac{12 H^2 \left(\frac{H}{M}\right)^{2 \lambda }}{M^2}\right)\right)}.\label{HdHumod2}
\end{eqnarray}
Note that when setting $\beta=0$ and $\alpha=1/(6M^{2})$ the result converts to that of Ref.\cite{Chatrabhuti:2015mws}. During inflation we can assume $H\simeq {\rm constant.}$, and then in this situation we obtain from Eq.(\ref{2sft})
\ba
H &\simeq& H_i +\frac{\left(\frac{H_i}{M}\right)^{-2 \lambda } \left(12 \beta  H^2_i \left(\frac{H_i}{M}\right)^{2 \lambda }-1\right)}{18 (\lambda +1) \left(2 \alpha +3 \beta +2 \beta  \log \left(\frac{12 H^2_i \left(\frac{H_i}{M}\right)^{2 \lambda }}{M^2}\right)\right)}(t-t_i)\,,\label{Htkmod2}
\ea
and 
\ba
a \simeq a_i \exp \Bigg\{H_i(t-t_i) +\frac{\left(\frac{H_i}{M}\right)^{-2 \lambda } \left(12 \beta  H^2_i \left(\frac{H_i}{M}\right)^{2 \lambda }-1\right)}{18 (\lambda +1) \left(2 \alpha +3 \beta +2 \beta  \log \left(\frac{12 H^2_i \left(\frac{H_i}{M}\right)^{2 \lambda }}{M^2}\right)\right)}\frac{(t-t_i)^{2}}{2}\Bigg\},
\ea
where $H_i$ and $a_i$ are respectively the Hubble parameter and the scale factor at the onset of inflation ($t=t_i$). The slow-roll parameter $\epsilon_1$ is defined by $\epsilon_1 \equiv -\dot{H}/H^2$ which in this case can be estimated as
\begin{eqnarray}
\epsilon_{1}\equiv -\frac{\dot{H}}{H^{2}}\simeq -\frac{\left(\frac{H}{M}\right)^{-2 \lambda } \left(12 \beta  H^2 \left(\frac{H}{M}\right)^{2 \lambda }-1\right)}{18 (\lambda +1)H^{2} \left(2 \alpha +3 \beta +2 \beta  \log \left(\frac{12 H^2 \left(\frac{H}{M}\right)^{2 \lambda }}{M^2}\right)\right)}.\label{epumod2}
\end{eqnarray}
We can check that $\epsilon_{1}$ is less than unity during inflation ($H^2 \gg M^2$) and we find when setting $n=2,\,\alpha=1/(6M^{2})$ that the bove expression reduces to $\epsilon_1 \simeq\frac{H^{-2 (\lambda +1)} M^{2 \lambda +2}}{6 (\lambda +1)}$. One can simply determine the time when inflation ends ($t=t_f$) by solving $\epsilon(t_f) \simeq 1$ to obtain
\ba
t_f &\simeq& t_i -\Bigg[\frac{\left(\frac{H_i}{M}\right)^{-2 \lambda } \left(12 \beta  H^2_i \left(\frac{H_i}{M}\right)^{2 \lambda }-1\right)}{18 (\lambda +1)H_i \left(2 \alpha +3 \beta +2 \beta  \log \left(\frac{12 H^2_i \left(\frac{H_i}{M}\right)^{2 \lambda }}{M^2}\right)\right)}\Bigg]^{-1}\, . \label{tfmod1}
\ea
The number of e-foldings from $t_i$ to $t_f$ is then given by
\ba
N &\equiv& \int^{t_f}_{t_i} Hdt \simeq H_i(t-t_i)+\frac{\left(\frac{H_i}{M}\right)^{-2 \lambda } \left(12 \beta  H^2_i \left(\frac{H_i}{M}\right)^{2 \lambda }-1\right)}{18 (\lambda +1)H^{2}_i \left(2 \alpha +3 \beta +2 \beta  \log \left(\frac{12 H^2_i \left(\frac{H_i}{M}\right)^{2 \lambda }}{M^2}\right)\right)}\frac{(t-t_i)^{2}}{2}\nonumber\\&\simeq& \frac{1}{2\epsilon_{1}(t_{i})}\, .
\ea
Note that when $\alpha=1/(6M^{2}),\,\beta=0$ and $\lambda=0$ the result is the same as that of the Starobinsky model.

\subsection{Model III: Einstein-Hu-Sawicki}
The Hu-Sawicki model of $f(R)$ gravity was initially proposed in Ref.\cite{Hu:2007nk}. It is a class of metric-variation $f(R)$ models that can describe the expansion of the universe without invoking a cosmological constant and satisfies both cosmological and solar-system tests in the small-field limit of the parameter space. The Einstein-Hu-Sawicki model of $f(R)$ is of the form:
\begin{eqnarray}
f(R)=R-M^{2}\frac{c_{1}\Big(R/M^{2}\Big)^{n}}{c_{2}\Big(R/M^{2}\Big)^{n}+1},
\end{eqnarray}
where $M$ is a mass scale, $c_{1},\,c_{2}$ and $n$ are arbitrary constants. The solar-system tests place constraints on these values. The authors of Ref.\cite{Hu:2016zrh} tested Hu-Sawicki $f(R)$ gravity using the effective field
theory approach and suggested that $c_{1}/c_{2}\approx 6\,\Omega_{\Lambda}/\Omega_{m}$ for $n=1,\,4$. However in the present work we consider inflationary model and instead keep $n$ fixed with $n=2$. Constants $c_{1},\,c_{2},\,M$ will be determined using latest inflationary constraints reported by PLANCK 2018 \cite{Akrami:2018odb}. From Eq.(\ref{Hdd}), we find for this model
\begin{eqnarray}
&&\Big(H^2 M^{12} \tilde{f}-2 M^{10} \big(2 H \dot{\tilde{f}}^3+\big(11 H^2+6 \dot{H}\big) \dot{\tilde{f}}^2 \tilde{f}+\big(11 H^3+13 H \dot{H}+\ddot{H}\big) \tilde{f}\big)\nonumber\\&&+2 \dot{\tilde{f}} \tilde{f} \big(H \ddot{\tilde{f}}+\big(9 H^3+10 H \dot{H}+\ddot{H}\big) \tilde{f}\big)+3 \tilde{f}^2 \big(2 H^2 \ddot{\tilde{f}}+\big(6 H^2 \dot{H}-\dot{H}^2+2 H \ddot{H}\big) \tilde{f}\big)\big) c_1\nonumber\\&&+108 M^6 \tilde{f}^2 \big(H \dot{\tilde{f}}+\big(2 H^2+\dot{H}\big) \tilde{f}\big)^2 \big(H^2 M^2 \tilde{f}+4 \big(H \dot{\tilde{f}}^3+\big(7 H^2+3 \dot{H}\big) \dot{\tilde{f}}^2 \tilde{f}\nonumber\\&&+\dot{\tilde{f}} \tilde{f} \big(H \ddot{\tilde{f}}+H \tilde{f}^2 \big(3 H \ddot{\tilde{f}}+\big(-2 H^3+11 H \dot{H}+3 \ddot{H}\big) \tilde{f}\big)\big) c_1\big) c_2\nonumber\\&&-3888 M^2 \tilde{f}^5 \big(H \dot{\tilde{f}}+\big(2 H^2+\dot{H}\big) \tilde{f}\big)^4 \big(-H^2 M^2+2 \big(H \dot{\tilde{f}}+\big(2 H^2+\dot{H}\big) \tilde{f}\big)^2 c_1\big) c_2^2\nonumber\\&&+46656 H^2 \tilde{f}^7 \big(H \dot{\tilde{f}}+\big(2 H^2+\dot{H}\big) \tilde{f}\big)^6 c_2^3\Big)=0\,,\label{hdd3}
\end{eqnarray}
and
\begin{eqnarray}
&&-2 c_1M^6 \dot{R} \tilde{f} \big(4 \dot{\tilde{f}}+9 H \tilde{f}\big)  \left(M^4-3 R^2 c_2\right) \left(M^4+R^2 c_2\right)+72 M^6 R \dot{R}^2 \tilde{f}^2 c_1 c_2 \left(M^4-R^2 c_2\right)\nonumber\\&&+R \left(M^4+R^2 c_2\right) \big(M^8+R^2 c_2 \left(2 M^4-2 M^2 R c_1+R^2 c_2\right)\big)\nonumber\\&&+\left(M^4+R^2 c_2\right) \big(-6 M^6 \ddot{R} \tilde{f}^2 c_1 \left(M^4-3 R^2 c_2\right)\big)=0\,.\label{Huff}
\end{eqnarray}
Here we are only interested in an inflationary solution. Therefore we invoke the slow-roll approximations. Hence the terms containing $\ddot{H}$ and higher power in $\dot{H}$ can be neglected in this particular regime. It is trivial to show that the Eq.(\ref{hdd3}) is reduced to
\begin{eqnarray}
\dot{H}&\simeq& -\frac{H^2}{3}
\Bigg(1+\lambda+\frac{2 (\lambda +1) M^4 \left(-6 c_{1} M^2 H^2 \left(\frac{H}{M}\right)^{2 \lambda }+144 c_{2} H^4 \left(\frac{H}{M}\right)^{4 \lambda }+M^4\right)}{288 c_{2} H^4 \left(\frac{H}{M}\right)^{4 \lambda } \left(-12 c_{1} M^2 H^2 \left(\frac{H}{M}\right)^{2 \lambda }+72 c_{2} H^4 \left(\frac{H}{M}\right)^{4 \lambda }+M^4\right)+M^8}\nonumber\\&&\quad\quad\quad-\frac{3(\lambda +1) M^4}{144 c_{2}H^4 \left(\frac{H}{M}\right)^{4\lambda}+M^4}\Bigg)^{-1}.\label{HdHu}
\end{eqnarray}
Setting $c_{1}=-1/6,\,c_{2}=0$, we obtain the same result given in Ref.\cite{Chatrabhuti:2015mws}. Moreover, setting both $c_{1},\,c_{2}$ and $\lambda=0$ to vanish, the result converts to the standard Starobinsky model \cite{Starobinsky:1980te}. During inflation we can assume $H\simeq {\rm constant.}$, and then in this situation we obtain from Eq.(\ref{HdHu})
\ba
H &\simeq& H_i -\frac{H_i^2}{3}
\Bigg(\frac{2 (\lambda +1) M^4 \left(-6 c_{1} M^2 H_i^2 \left(\frac{H_i}{M}\right)^{2 \lambda }+144 c_{2} H_i^4 \left(\frac{H_i}{M}\right)^{4 \lambda }+M^4\right)}{288 c_{2} H_i^4 \left(\frac{H_i}{M}\right)^{4 \lambda } \left(-12 c_{1} M^2 H_i^2 \left(\frac{H_i}{M}\right)^{2 \lambda }+72 c_{2} H_i^4 \left(\frac{H_i}{M}\right)^{4 \lambda }+M^4\right)+M^8}\nonumber\\&&\quad\quad\quad\quad\quad+1+\lambda-\frac{3(\lambda +1) M^4}{144 c_{2}H_i^4 \left(\frac{H_i}{M}\right)^{4\lambda}+M^4}\Bigg)^{-1}(t-t_i)\,,\label{Htk}
\ea
and 
\ba
a &\simeq& a_i \exp \Bigg\{H_i(t-t_i) -\frac{H_i^2}{6}
\Bigg(\frac{2 (\lambda +1) M^4 \left(-6 c_{1} M^2 H_i^2 \left(\frac{H_i}{M}\right)^{2 \lambda }+144 c_{2} H_i^4 \left(\frac{H_i}{M}\right)^{4 \lambda }+M^4\right)}{288 c_{2} H_i^4 \left(\frac{H_i}{M}\right)^{4 \lambda } \left(-12 c_{1} M^2 H_i^2 \left(\frac{H_i}{M}\right)^{2 \lambda }+72 c_{2} H_i^4 \left(\frac{H_i}{M}\right)^{4 \lambda }+M^4\right)+M^8}\nonumber\\&&\quad\quad\quad\quad\quad+1+\lambda-\frac{3(\lambda +1) M^4}{144 c_{2}H_i^4 \left(\frac{H_i}{M}\right)^{4\lambda}+M^4}\Bigg)^{-1}\frac{(t-t_i)^{2}}{2}\Bigg\},
\ea
where $H_i$ and $a_i$ are respectively the Hubble parameter and the scale factor at the onset of inflation ($t=t_i$). The slow-roll parameter $\epsilon_1$ is defined by $\epsilon_1 \equiv -\dot{H}/H^2$ which in this case can be estimated as
\begin{eqnarray}
\epsilon_{1}\equiv -\frac{\dot{H}}{H^{2}}&\simeq& \frac{1}{3}
\Bigg(1+\lambda+\frac{2 (\lambda +1) M^4 \left(-6 c_{1} M^2 H^2 \left(\frac{H}{M}\right)^{2 \lambda }+144 c_{2} H^4 \left(\frac{H}{M}\right)^{4 \lambda }+M^4\right)}{288 c_{2} H^4 \left(\frac{H}{M}\right)^{4 \lambda } \left(-12 c_{1} M^2 H^2 \left(\frac{H}{M}\right)^{2 \lambda }+72 c_{2} H^4 \left(\frac{H}{M}\right)^{4 \lambda }+M^4\right)+M^8}\nonumber\\&&-\frac{3(\lambda +1) M^4}{144 c_{2}H^4 \left(\frac{H}{M}\right)^{4\lambda}+M^4}\Bigg)^{-1}.\label{epu}
\end{eqnarray}
We can check that $\epsilon_{1}$ is less than unity during inflation ($H^2 \gg M^2$) and we find when setting $c_{1}=-1/6,\,c_{2}=0$ that the above expression reduces to $\epsilon_1 \simeq\frac{H^{-2 (\lambda +1)} M^{2 \lambda +2}}{6 (\lambda +1)}$. One can simply determine the time when inflation ends ($t=t_f$) by solving $\epsilon(t_f) \simeq 1$ to obtain
\ba
t_f &\simeq& t_i +\frac{3}{H_i}
\Bigg(\frac{2 (\lambda +1) M^4 \left(-6 c_{1} M^2 H_i^2 \left(\frac{H_i}{M}\right)^{2 \lambda }+144 c_{2} H_i^4 \left(\frac{H_i}{M}\right)^{4 \lambda }+M^4\right)}{288 c_{2} H_i^4 \left(\frac{H_i}{M}\right)^{4 \lambda } \left(-12 c_{1} M^2 H_i^2 \left(\frac{H_i}{M}\right)^{2 \lambda }+72 c_{2} H_i^4 \left(\frac{H_i}{M}\right)^{4 \lambda }+M^4\right)+M^8}\nonumber\\&&\quad\quad\quad+1+\lambda-\frac{3(\lambda +1) M^4}{144 c_{2}H_i^4 \left(\frac{H_i}{M}\right)^{4\lambda}+M^4}\Bigg)\, . \label{tf}
\ea
The number of e-foldings from $t_i$ to $t_f$ is then given by
\ba
N &\equiv& \int^{t_f}_{t_i} Hdt \simeq H_i(t_{f}-t_i) \nonumber\\&&-\frac{H_i^2}{6}
\Bigg(\frac{2 (\lambda +1) M^4 \left(-6 c_{1} M^2 H_i^2 \left(\frac{H_i}{M}\right)^{2 \lambda }+144 c_{2} H_i^4 \left(\frac{H_i}{M}\right)^{4 \lambda }+M^4\right)}{288 c_{2} H_i^4 \left(\frac{H_i}{M}\right)^{4 \lambda } \left(-12 c_{1} M^2 H_i^2 \left(\frac{H_i}{M}\right)^{2 \lambda }+72 c_{2} H_i^4 \left(\frac{H_i}{M}\right)^{4 \lambda }+M^4\right)+M^8}\nonumber\\&&+1+\lambda-\frac{3(\lambda +1) M^4}{144 c_{2}H_i^4 \left(\frac{H_i}{M}\right)^{4\lambda}+M^4}\Bigg)^{-1}\frac{(t-t_i)^{2}}{2} \nonumber\\&\simeq& \frac{1}{2\epsilon_{1}(t_{i})}\,.\label{epN3}
\ea
Note that when $c_{1}=-1/6,\,c_{2}=0$ and $\lambda=0$ the result is the same as that of the Starobinsky model.
%%%%%%%%%%%%%%%%%%%%%%%%%%%%%%%%%%%%%%%%%%
\section{A short review of cosmological perturbation in
$f(R)$ gravity's rainbow}\label{sec3}
%%%%%%%%%%%%%%%%%%%%%%%%%%%%%%%%%%%%%%%
In this section, we take a short review of the cosmological perturbation in $f(R)$ gravity in the present of the gravity's rainbow effect. We will divide this part into three subsections, i.e. perturbation equations, the curvature perturbation and the tensor perturbation.
\subsection{Perturbation Equations}
For $\tilde{g}(\epsilon )=1$, the general perturbed metric of a flat FLRW metric with gravity's rainbow effect is given by \cite{Chatrabhuti:2015mws} 
\begin{eqnarray}\label{a1}
    ds^2 &=&-\frac{(1+2\alpha )}{\tilde{f}^2(\epsilon)}dt^2-\frac{2 a(t) \left(\partial_i\beta-S_i\right)}{\tilde{f}(\epsilon )}dt dx^i\nonumber\\&&+a^2(t)  \left(\delta _{ij}+2 \psi  \delta _{ij}+2 \partial_i\partial_j \gamma +2\partial_j F_i+h_{ij}\right)dx^i dx^j\,,
\end{eqnarray}
where $\alpha,\beta,\psi,\gamma$ are scalar perturbations, $S_i,F_i$ are vector perturbations and $h_ij$ are tensor perturbations. For our purpose, we consider the scalar perturbations and tensor perturbations separately and ignore the vector perturbations, i.e. $S_i=F_i=0$. Then, let us consider the gauge transformation containing the gravity's rainbow effect as follows \cite{Chatrabhuti:2015mws}:
\begin{eqnarray}
   \alpha \to \hat{\alpha }&=&\alpha+\frac{ \dot{\tilde{f}}}{\tilde{f}}\delta t -\dot{\delta } t\\
   \beta \to \hat{\beta }&=&\beta-\frac{\delta t}{a \tilde{f}}+a \tilde{f}\dot{\delta }x \\
   \psi \to \hat{\psi }&=&\psi -H\delta t \\
   \gamma \to \hat{\gamma }&=&\gamma -\delta x.
\end{eqnarray}
Note that the tensor perturbation $h_{ij}$ are invariant under this gauge transformation. Thus, gauge invariant quantities according to gauge transformation given above read
\begin{eqnarray}
    \Phi &=&\alpha -\tilde{f}\frac{d}{dt}\left[a^{2}\tilde{f} \left(\dot{\gamma} + \frac{\beta}{a\tilde{f}}\right)\right]\,,\\
    \Psi &=&-\psi+a^2\tilde{f}^2 H\left(\dot{\gamma} + \frac{\beta}{a\tilde{f}}\right)\,,\\
    \mathcal{R}&=&\psi -\frac{ H\delta F}{\dot{F}}.
\end{eqnarray}
We can choose $\beta=0$ and $\gamma=0$, then, $\Phi=\alpha$ and $\Psi=-\psi$. Therefore, the metric (\ref{a1}) becomes
\begin{eqnarray}\label{a9}
    ds^2 = -\frac{1+2\Phi}{\tilde{f}^2(t)}dt^2 + a^2(t)(1-2\Psi)\delta_{ij} dx^i dx^j\,.
\end{eqnarray}
For simplicity, we define a new variable $A \equiv 3(H\Phi+\dot{\Psi})$. With the metric (\ref{a9}) and Eq.(\ref{eom}), we obtain the following system of equations
\begin{eqnarray}
-\frac{\nabla^2\Psi}{a^2}+\tilde{f}^2HA &=& -\frac{1}{2F}\left[3\tilde{f}^2\left(H^2 + \dot{H} + \frac{\dot{\tilde{f}}}{\tilde{f}}\right)\delta F + \frac{\nabla^2\delta F}{a^2}-3\tilde{f}^2H\delta\dot{F} \right. \nonumber \\ 
 &&+ \left. 3\tilde{f}^2H\dot{F}\Phi + \tilde{f}^2\dot{F}A+\kappa^2\delta\rho\right] \, , \label{a11} \\
H\Phi+\dot{\Psi}&=&-\frac{1}{2F}(H\delta F+\dot{F}\Phi-\delta\dot{F}) \, ,  \label{a12} 
\end{eqnarray}
and 
\begin{eqnarray}
\dot{A} &+& \left(2H+\frac{\dot{\tilde{f}}}{\tilde{f}}\right)A+3\dot{H}\Phi + \frac{\nabla^2\Phi}{a^2\tilde{f}^2}+\frac{3H\Phi\dot{\tilde{f}}}{\tilde{f}} 
= \frac{1}{2F}\left[3\delta\ddot{F}+3\left(H+\frac{\dot{\tilde{f}}}{\tilde{f}}\right)\delta\dot{F} \right. \nonumber \\ &-&\left. 6H^2\delta F -\frac{\nabla^2\delta F}{a^2\tilde{f}^2} - 3\dot{F}\dot{\Phi}-\dot{F}A - 3\left(H+\frac{\dot{\tilde{f}}}{\tilde{f}}\right)\dot{F}\Phi-6\ddot{F}\Phi+\frac{\kappa^2}{\tilde{f}^2}(3\delta P_+\delta\rho) \right] \, .
\label{a13}
\end{eqnarray}
We will use these equations to study the scalar perturbations during inflation. Thus, we will not include a perfect fluid into our consideration, i.e. $\delta \rho=0$ and $\delta P=0$.

\subsection{Curvature Perturbation}
This subsection deals with the scalar perturbation generated during inflation. For the case of $\delta F=0$, hence, the curvature perturbation on a constant-time hypersurface implies $\mathcal{R}=\phi =-\Psi$. Thus, Eq.(\ref{a12}) becomes
\begin{eqnarray}\label{a14}
 \Phi = \frac{\dot{\mathcal{R}}}{H+\dot{F}/2F}\,.
\end{eqnarray}
Then, inserting Eq.(\ref{a14}) into equation (\ref{a11}) leads to
\begin{eqnarray}
 A = -\frac{1}{H+\dot{F}/2F}\left[\frac{\nabla^2\mathcal{R}}{a^2\tilde{f}^2}+\frac{3H\dot{F}\dot{\mathcal{R}}}{2F(H+\dot{F}/2F)}\right] \ . \label{a15}
\end{eqnarray}
 By using the background equation (\ref{ijcom}) and (\ref{a13}), we find
 \begin{eqnarray}
 \dot{A}+\left(2H+\frac{\dot{F}}{2F}\right)A+\frac{\dot{\tilde{f}}A}{\tilde{f}}+\frac{3\dot{F}\dot{\Phi}}{2F}+\left[\frac{3\ddot{F}+6H\dot{F}}{2F}+\frac{\nabla^2}{a^2\tilde{f}^2}\right]\Phi+\frac{3\dot{F}}{2F}\frac{\Phi\dot{\tilde{f}}}{\tilde{f}} = 0. \label{a16}
\end{eqnarray}
Substituting Eqs.(\ref{a14}) and (\ref{a15}) into (\ref{a16}) allows us to write the equation of $\mathcal{R}$ in a Fourier space as
\begin{eqnarray}
 \ddot{\mathcal{R}} + \frac{1}{a^3Q_s}\frac{d}{dt}(a^3Q_s)\dot{\mathcal{R}} + \frac{\dot{\tilde{f}}}{\tilde{f}}\dot{\mathcal{R}} + \frac{k^2}{a^2\tilde{f}^2} \mathcal{R}= 0 \ , \label{a17}
\end{eqnarray}
where $k$ is a comoving wave number and a new variable $Q_s$ is defined by
\begin{eqnarray}
 Q_s \equiv \frac{3\dot{F}^2}{2\kappa^2F(H+\dot{F}/2F)^2} \ . \label{a18}
\end{eqnarray}
Eq.(\ref{a17}) can be reduced to 
\begin{eqnarray}
 u'' + \left(k^2-\frac{z_s''}{z_s}\right)u = 0 \ , \label{a19}
\end{eqnarray}
where new parameters $z_s = a\sqrt{Q_s}$, $u = z_s\mathcal{R}$ and a prime denotes a derivative with respect to the new time coordinates $\eta = \int (a\tilde{f})^{-1} dt$. In order to determine the spectrum of curvature perturbations we define various slow-roll parameters as
\begin{eqnarray}
 \epsilon_1 \equiv -\frac{\dot{H}}{H^2},  \ \ \epsilon_2 \equiv \frac{\dot{F}}{2HF}, \ \ \epsilon_3 \equiv \frac{\dot{E}}{2HE}\ ,
\end{eqnarray}\label{a20}
where $E \equiv 3\dot{F}^2/2\kappa^2$.   Subsequently, $Q_s$ can be rewritten to obtain
\begin{eqnarray}
 Q_s = \frac{E}{FH^2(1+\epsilon_2)^2} \ . \label{a22}
\end{eqnarray}
During inflationary era, parameters $\epsilon_i$ are assumed to be constant $(\dot{\epsilon}_i\simeq0)$ and in this work we assume that $\tilde{f} =1+(H/M)^{\lambda}$. Hence, we are able to determine $\eta$: 
\begin{align}
    \eta = -\frac{1}{(1-(1+\lambda)\epsilon_1)\tilde{f}aH}\,.
\end{align}
 The term $z_s''/z_s$ in equation (\ref{a19}) can be approximated to yield
 \begin{eqnarray}\label{a24}
 \frac{z_s''}{z_s} = \frac{\nu^2_{\mathcal{R}} - 1/4}{\eta^2} \ , 
\end{eqnarray}
with
\begin{eqnarray}\label{a25}
 \nu_{\mathcal{R}}^2 = \frac{1}{4} + \frac{(1+\epsilon_1 - \epsilon_2+\epsilon_3)(2-\lambda\epsilon_1 -\epsilon_2+\epsilon_3)}{(1-(\lambda+1)\epsilon_1)^2} \ .
\end{eqnarray}
Therefore, the estimated solution of Eq.(\ref{a19}) can be written in terms of a linear combination of Hankel functions
\begin{eqnarray}
 u = \frac{\sqrt{\pi|\eta|}}{2}\textmd{e}^{i(1+2\nu_{\mathcal{R}})\pi/4}\left[b_1\textmd{H}_{\nu_{\mathcal{R}}}^{(1)}(k|\eta|)+b_2\textmd{H}_{\nu_{\mathcal{R}}}^{(2)}(k|\eta|)\right] \ , \label{a26}
\end{eqnarray}
where $b_1$, $b_2$ are integration constants and $\textmd{H}_{\nu_{\mathcal{R}}}^{(1)}(k|\eta|)$, $\textmd{H}_{\nu_{\mathcal{R}}}^{(2)}(k|\eta|)$ are the Hankel functions of the first kind and the second kind, respectively.
In the asymptotic past $k\eta \rightarrow -\infty$, the estimated solution (\ref{a26}) will become $u \rightarrow \textmd{e}^{-ik\eta}/\sqrt{2k}$. This tells us that $b_1=1$ and $b_2=0$. Thus the estimated solution can be expressed as
\begin{eqnarray}
 u = \frac{\sqrt{\pi|\eta|}}{2}\textmd{e}^{i(1+2\nu_{\mathcal{R}})\pi/4}\textmd{H}_{\nu_{\mathcal{R}}}^{(1)}(k|\eta|) \ . \label{a27}
\end{eqnarray}
Using the definition of the power spectrum of curvature perturbations as
\begin{eqnarray}
 \mathcal{P}_{\mathcal{R}} \equiv \frac{4\pi k^3}{(2\pi)^3}|\mathcal{R}|^2 \ , \label{a28}
\end{eqnarray}
together with the estimated solution (\ref{a27}) and $u = z_s\mathcal{R}$, we obtain
\begin{eqnarray}
 \mathcal{P}_{\mathcal{R}} = \frac{1}{Q_s}\left[(1-(1+\lambda)\epsilon_1)\frac{\Gamma(\nu_{\mathcal{R}})H}{2\pi\Gamma(3/2)}\left(\frac{H}{M}\right)^\lambda\right]^2\left(\frac{k|\eta|}{2}\right)^{3-2\nu_{\mathcal{R}}} \ , \label{a29}
\end{eqnarray}
where we have used $\textmd{H}_{\nu_{\mathcal{R}}}^{(1)}(k|\eta|) \rightarrow -(i/\pi)\Gamma(\nu_{\mathcal{R}})(k|\eta|/2)^{-\nu_{\mathcal{R}}}$ for $k|\eta| \rightarrow 0$. Since $\mathcal{R}$ is fixed after the Hubble radius crossing, $P_{\mathcal{R}}$ should be evaluated at $k=aH$. Instantly, we define the spectral index $n_{\mathcal{R}}$  as
\begin{eqnarray}
 n_{\mathcal{R}} - 1 = \left.\frac{d\textmd{ln}\mathcal{P}_{\mathcal{R}}}{d\textmd{ln}k}\right|_{k=aH} = 3 - 2\nu_{\mathcal{R}} \ . \label{a30}
\end{eqnarray}
Consequently, the spectral index can be written in terms of the slow-roll parameters as
\begin{eqnarray}
 n_{\mathcal{R}} - 1 \simeq -2(\lambda+2)\epsilon_1+2\epsilon_2-2\epsilon_3 \ , \label{a31}
\end{eqnarray}
where during the inflationary epoch, we have assumed that $|\epsilon_i | \ll 1$. Notice that the spectrum is nearly scale-invariant when $|\epsilon_i|$ are much smaller than unity, i.e. $n_{\mathcal{R}} \simeq 1$. Subsequently, the power spectrum of curvature perturbation takes the form
\begin{eqnarray}
 \mathcal{P}_{\mathcal{R}} \approx \frac{1}{Q_s}\left(\frac{H}{2\pi}\right)^2\left(\frac{H}{M}\right)^{2\lambda} \ . \label{a32}
\end{eqnarray}
In Sec.\ref{sec4}, we can constrain parameters for the $f(R)$ models using Eq.(\ref{a32}) since $\mathcal{P}_{\mathcal{R}}$ contains $Q_s$ which is a function of $F(R)$.
%%%%%%%%%%%%%%%%%%%%%%%%%%%%%%%%%
\subsection{Tensor Perturbation}
In this subsection, we explore how to derive the power spectrum and spectral index of the tensor perturbation. Generally, the tensor perturbation $h_{ij}$ are written as
\begin{eqnarray}
 h_{ij} = h_{+}e^+_{ij} + h_{\times}e^\times_{ij} \ , \label{a33}
\end{eqnarray}
where $e^+_{ij}$ and $e^\times_{ij}$ are the polarization tensors corresponding to the two polarization states of $h_{ij}$. Suppose that $\vec{k}$ is in the direction along the $z$-axis, then the non-vanishing components of polarization tensors are $e^+_{xx} = -e^+_{yy} = 1$ and $e^\times_{xy} = e^\times_{yx} = 1$. With only tenser perturbation, the perturbed FLRW metric (\ref{a1}) can be written as
\begin{eqnarray}
 ds^2 = -\frac{dt^2}{\tilde{f}(\varepsilon)^2} + a^2(t)h_{\times}dxdy + a^2(t)\left[(1+h_{+})dx^2+(1-h_{+})dy^2+dz^2\right] .\label{34}
\end{eqnarray}
Using this above metric in equation (\ref{eom}), we can show that the Fourier components $h_\chi$ yields the following equation
\begin{eqnarray}
  \ddot{h}_\chi + \frac{(a^3F)^\cdot}{a^3F}\dot{h}_\chi + \frac{\dot{\tilde{f}}}{\tilde{f}}\dot{h}_\chi + \frac{k^2}{a^2\tilde{f}^2}h_\chi = 0 \ , \label{a35}
\end{eqnarray}
where $\chi$ denotes polarizations $+$ and $\times$. At this point, we have proceeded using the procedure similar to the case of curvature perturbation and introduced the new variables $z_t = a\sqrt{F}$ and $u_\chi = z_t h_\chi /\sqrt{2 \kappa^{2}}$. Therefore Eq.(\ref{a35}) can be rewritten as
\begin{eqnarray}
 u''_\chi + \left(k^2-\frac{z_t''}{z_t}\right)u_\chi = 0 \ . \label{a36}
\end{eqnarray}
Notice that for a massless scalar field $u_\chi$ has dimension of mass. Assuming $\dot{\epsilon}_i=0$, we obtain
\begin{eqnarray}
 \frac{z_t''}{z_t} = \frac{\nu^2_t-1/4}{\eta^2} \ , \label{a37}
\end{eqnarray}
where
\begin{eqnarray}
 \nu^2_t = \frac{1}{4} + \frac{(1+\epsilon_2)(2-(1+\lambda)\epsilon_1+\epsilon_2)}{(1-(1+\lambda)\epsilon_1)^2} \ . \label{a38}
\end{eqnarray}
Alike curvature perturbation, the estimated solution to Eq.(\ref{a36}) can be also expressed in terms of a linear combination of Hankel functions. Thus, the power spectrum of tensor perturbations $P_T$ after the Hubble radius crossing can be estimated as 
\begin{eqnarray}
 \mathcal{P}_T &=& 4\times\frac{2\kappa^{2}}{a^2F}\frac{4\pi k^3}{(2\pi)^3}|u_\chi|^2  \nonumber\\&=& \frac{16}{\pi}\left(\frac{H}{M_{P}}\right)^2\frac{1}{F}\left[(1-(1+\lambda)\epsilon_1)\frac{\Gamma(\nu_t)}{\Gamma(3/2)}\left(\frac{H}{M}\right)^\lambda\right]^2\left(\frac{k|\eta|}{2} \right)^{3-2\nu_t},  
 \label{a39}
\end{eqnarray}
where we have used $\tilde{f} \simeq (H/M)^{\lambda}$ and $\nu_t$ can be obtained by assuming that the slow-roll parameters are very small during inflation to obtain
\begin{eqnarray}
 \nu_t \simeq \frac{3}{2} + (1+\lambda)\epsilon_1 + \epsilon_2 \ . \label{a40}
\end{eqnarray}
Additionally, the spectral index of tensor perturbations is obtained via
\begin{eqnarray}
 n_T = \left.\frac{d \textmd{ln}\mathcal{P}_T}{d\textmd{ln}k}\right|_{k=aH} = 3-2\nu_t \simeq -2(1+\lambda)\epsilon_1 - 2\epsilon_2 \ . \label{a41}
\end{eqnarray}
The power spectrum $\mathcal{P}_T$ can also be given by
\begin{eqnarray}
 \mathcal{P}_T \simeq \frac{16}{\pi}\left(\frac{H}{M_{P}}\right)^2\frac{1}{F}\left(\frac{H}{M}\right)^{2\lambda} \ . \label{a42}
\end{eqnarray}
Also, the tensor-to-scalar ratio $r$ can be determined by invoking the following definition:
\begin{eqnarray}
 r \equiv \frac{\mathcal{P}_T}{\mathcal{P}_R}  \simeq \frac{64\pi}{M_{P}^2}\frac{Q_s}{F} \ . \label{a43}
\end{eqnarray}
Substituting $Q_s$ from Eq.(\ref{a18}), we finally obtain
\begin{eqnarray}
 r = 48\epsilon_2^2 \ . \label{a44}
\end{eqnarray}
In the next section, we consider the spectra of perturbations based on various $f(R)$ models in gravity’s rainbow theory and confront the results predicted by our models with Planck 2018 data. It is worth noting that the current observational limit on the tensor-to-scalar ratio is $r<0.1$ \cite{Ade:2015lrj,Akrami:2018odb}. Some proposed experiments, such as CMBPol \cite{Andre:2013nfa}, PRISM \cite{Baumann:2008aq} and CORE \cite{Finelli:2016cyd}, can reach the $10^{-3}$ level. However it is expected that measuring $r<10^{-4}$ via CMB polarisation is extremely challenging (see e.g., Ref.\cite{Verde:2005ff}).

%%%%%%%%%%%%%%%%%%%%%%%%%%%%%%%%%%%%%%%%%%%%%%%%%%
\section{Confrontation with the Planck 2018 data}\label{sec4}
%%%%%%%%%%%%%%%%%%%%%%%%%%%%%%%%%%%%%%%%%%%%%%%%
In this section, we consider the scalar and tensor perturbation based on the Hu-Sawicki model with gravity's rainbow effect. Recall the definitions of slow-roll parameters in Sec.\ref{sec3}:
\begin{eqnarray}
 \epsilon_1 \equiv -\frac{\dot{H}}{H^2},  \ \ \epsilon_2 \equiv \frac{\dot{F}}{2HF}, \ \ \epsilon_3 \equiv \frac{\dot{E}}{2HE}=\frac{\ddot{F}}{H \dot{F}}\,
\end{eqnarray}\label{4.25}
where $E \equiv 3\dot{F}^2/2\kappa^2$.
From Eq.(\ref{a22}), $Q_s$ can be given by
\begin{eqnarray}
 Q_s = \frac{E}{FH^2(1+\epsilon_2)^2} \ . \label{4.27}
\end{eqnarray}
In order to derive the power spectra and the spectral indices, the relations between slow-roll parameters must be first verified. To this end, we recall the background equation (\ref{ijcom}) for $\rho=0=P$ and $\tilde{g}(\epsilon )=1$:
\begin{eqnarray}
    \ddot{F}-H\dot{F} +2 F \dot{H}+2 F H\frac{\dot{\tilde{f}}}{\tilde{f}}=0.\label{4.28}
\end{eqnarray}
By using $\tilde{f}\approx (H/M)^\lambda$, we have
\begin{eqnarray}
    \frac{\dot{\tilde{f}}}{\tilde{f}}=\frac{\lambda \dot{H}}{H}.
\end{eqnarray}
Thus, Eq.(\ref{4.28}) can be rewritten as
\begin{eqnarray}
    \ddot{F}-H\dot{F} +2(1+\lambda) F \dot{H}=0.\label{4.30}
\end{eqnarray}
Dividing Eq.(\ref{4.30}) by $2H^2F$, then it reduces to
\begin{eqnarray}
    \frac{\ddot{F}}{2H^2F}-\epsilon_2 -(1+\lambda)\epsilon_1=0.\label{4.31x}
\end{eqnarray}
Notice that we can use the relation (\ref{4.25}) to rewrite the above equation to yield
\begin{eqnarray}
    \epsilon_2(\epsilon_3-1)-(1+\lambda)\epsilon_1=0.\label{4.31}
\end{eqnarray}
The higher order powers of $\epsilon_i$ can be neglected since $\epsilon_i$, for all $i$, are very small during inflation era. Hence, we find
\begin{eqnarray}
    \epsilon_2 \simeq -(1+\lambda)\epsilon_1 . \label{ep1ep2}
\end{eqnarray}
We can verify another relation among slow-roll parameters by considering the definition of $\epsilon_3$
\begin{eqnarray}
    \epsilon_3 \equiv \frac{\dot{E}}{2HE}=\frac{\ddot{F}}{H \dot{F}}.
\end{eqnarray}
In order to verify the relations among slow-roll parameters, we will focus on some different forms of $f(R)$ given below.

\subsection{Model I: $f(R)=R+\alpha (R/M)^{n}$}
It is convenient to redefine a parameter $\alpha$ given in Eq.(\ref{star}) so that it becomes a dimensionless parameter. Here we take $\alpha\rightarrow\alpha M^{2}$ and then a function $f(R)$ becomes
\begin{eqnarray}
f(R)=R+\alpha M^{2} \Big(\frac{R}{M^{2}}\Big)^{n}. \label{starre}
\end{eqnarray}
Note that a re-definition in the present analysis does not affect our discussions in Sec.\ref{21}. Using Eq.(\ref{starre}), we can approximate $F(R)$ to obtain
\begin{eqnarray}
F(R)\approx 12^{-1+n}n \alpha \left(\left(\frac{H}{M}\right)^{2+2 \lambda }\right)^{-1+n},\,\label{FR1}
\end{eqnarray}
where the approximation is valid only when $H\gg M$ is assumed during inflation. We consider Eq.(\ref{a32}) and then the power spectrum of curvature perturbation reads
\begin{eqnarray}
    \mathcal{P}_{\mathcal{R}} \approx \frac{1}{Q_s}\left(\frac{H}{2\pi}\right)^2\left(\frac{H}{M}\right)^{2\lambda} = \frac{1}{3\pi F}\left(\frac{H}{m_{\rm P}}\right)^{2}\left(\frac{H}{M}\right)^{2\lambda}\frac{1}{(1+\lambda)^{2}\epsilon^{2}_{1}}\,.\label{PRmod1}
\end{eqnarray}
\begin{figure}[!h]	
	\includegraphics[width=8.2cm]{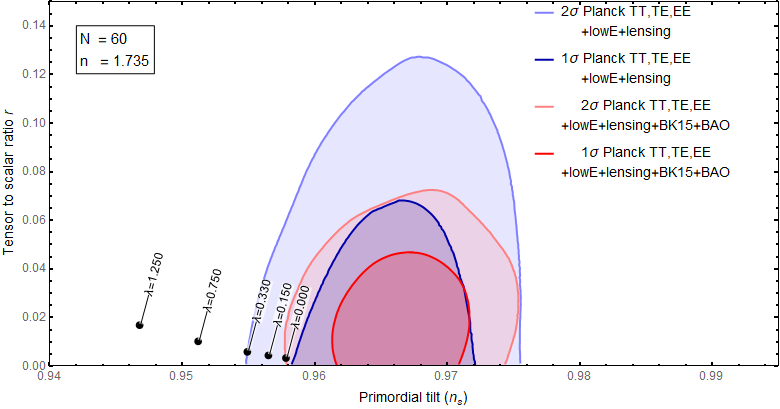}
	\includegraphics[width=8.2cm]{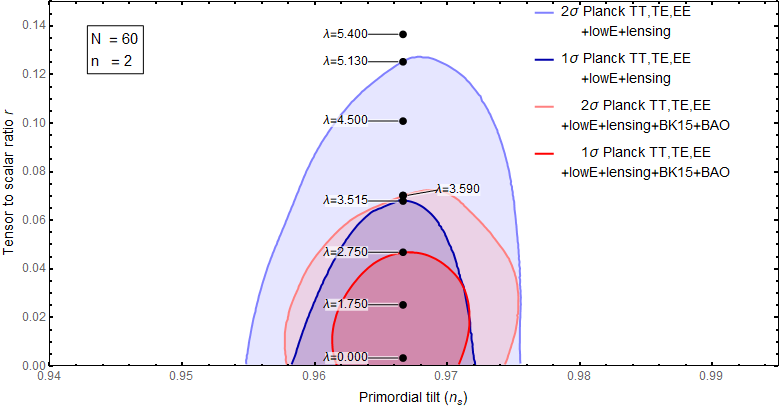}
	\includegraphics[width=8.2cm]{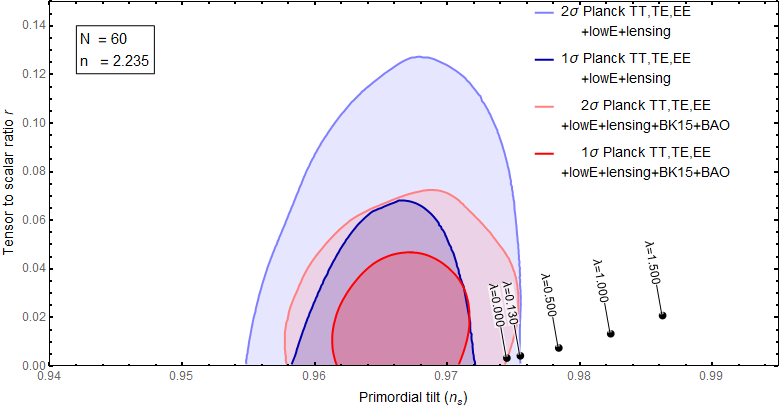}
	\includegraphics[width=8.2cm]{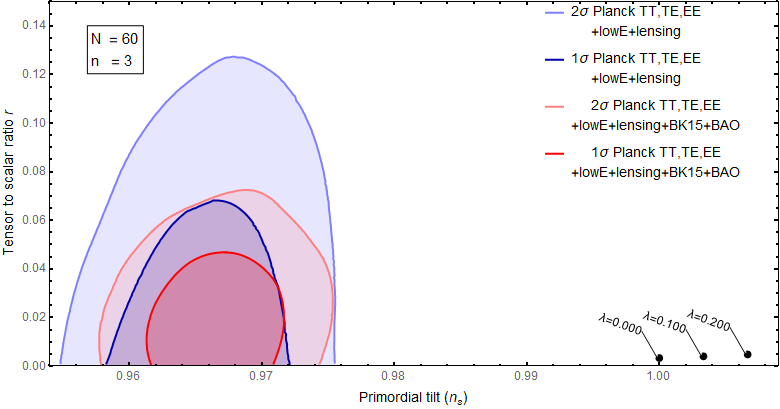}
	\centering
	\caption{We compare the theoretical predictions in the $(r-n_{s})$ plane for different values of $\lambda$ using $N=60$ and varying $n$. We plots for $N=60,\,n=1.735$ (upper-left panel); $N=60,\,n=2$ (upper-right panel); and for $N=60,\,n=2.235$ (lower-left panel); $N=60,\,n=3$ (lower-right panel) with Planck’15 results for TT, TE, EE, +lowE+lensing and +BK15+BAO.}
	\label{nsrmod1}
\end{figure}
We can further simplify the above result to yield
\begin{eqnarray}
    \mathcal{P}_{\mathcal{R}} \approx \frac{3^{-n} 4^{2-n} \left(\left(\frac{H}{M}\right)^{2+2 \lambda }\right)^{2-n} M^2}{4n \pi\alpha(1+\lambda )^2 m_{\rm P}^2}\frac{1}{(1+\lambda)^{2}\epsilon^{2}_{1}}\,,
\end{eqnarray}
when we have inserted $F(R)$ given in Eq.(\ref{FR1}) into Eq.(\ref{PRmod1}). It is worth noting that when $n=2$, we obtain a special case for which the Starobinky model is recovered. In this model, therefore, $\epsilon_3$ reads
\begin{eqnarray}
    \epsilon_3=\frac{\dot{H} (-2 \lambda +2 (\lambda +1) n-3)}{H^2}+\frac{\ddot{H}}{H \dot{H}}\,.\label{ep3}
\end{eqnarray}
Assuming slow-roll approximations, the terms containing $\ddot{H}$ can be ignored and then the relation between $\epsilon_3$ and $\epsilon_1$ reads
\begin{eqnarray}
    \epsilon_3=\epsilon _1 (2 \lambda -2 (\lambda +1) n+3)\,,\label{ep3mod1}
\end{eqnarray}
where we have used Eq.(\ref{ep1ep2}) and Eq.(\ref{ep3mod1}) together with Eq.(\ref{a31}). Hence, we have
\begin{eqnarray}
 n_{\mathcal{R}} - 1 &\simeq& -2(\lambda+2)\epsilon_1+2\epsilon_2-2\epsilon_3 \ ,\\
 n_{\mathcal{R}}&=&1+4 \epsilon _1 (\lambda  (n-2)+n-3)\ ,
\end{eqnarray}
where we have defined a new parameter $\Delta \equiv H/M$. For simplicity, let us suppose that during inflation the expansion is de
Sitter (exponential) with a constant Hubble parameter. In terms of the number of efoldings, $\mathcal{P}_{\mathcal{R}},\,n_\mathcal{R}$ and $r$ read
\begin{eqnarray}
    \mathcal{P}_{\mathcal{R}} &\approx& \frac{3^{-n} 4^{2-n} \left(\Delta^{2+2 \lambda }\right)^{2-n} M^2}{n \pi\alpha(1+\lambda )^2 m_{\rm P}^2}\frac{N^{2}}{(1+\lambda)^{2}},\label{PR33m1}\\
    n_\mathcal{R}&\approx&1-\frac{6}{N}+\frac{2 n}{N}-\frac{4 \lambda }{N}+\frac{2 \lambda  n}{N}, \label{finalnsm1} \\
    r&\approx& \frac{12 (\lambda +1)^2}{N^2}. \label{finalrm1}
\end{eqnarray}
We find that the above parameters reduce to those of the Starobinsky model when $n=2,\,\lambda=0,\,\alpha=1/6$. We now compare our predicted results with Planck 2018 data. We find from Fig.(\ref{nsrmod1}) for $N=60,\,n=1.735$ that the predictions are consistent with the Planck’15 results for TT, TE, EE, +lowE+lensing at two sigma confidence level for $0.000\leq\lambda\leq 0.330$ and lie outside the Planck’15 results for TT, TE, EE, +lowE+lensing+BK15+BAO for all values of $\lambda$. For $N=60,\,n=2$, our results are in excellent agreement with Planck’15 results for TT, TE, EE, +lowE+lensing and for TT, TE, EE, +lowE+lensing+BK15+BAO at one sigma confident level when $0.00\leq\lambda \leq 3.515$ and $0.00\leq\lambda \leq 2.750$, respectively.

Additionally, for $N=60$ and $n=2.235$, we discover that the predictions are consistent with the Planck’15 results for TT, TE, EE, +lowE+lensing at two sigma confidence level when $0.000\leq\lambda\leq 0.130$ and lie at the boundary of the two sigma confident level when $\lambda = 0.00$. However, for $N=60$ and $n=3$, our results are inconsistent with Planck’15 results for TT, TE, EE, +lowE+lensing and TT, TE, EE, +lowE+lensing+BK15+BAO. Using Planck 2018 data for $\mathcal{P}_{\mathcal{R}}$, we can solve for $n$ to obtain
\begin{eqnarray}
n &=& \frac{4.76\times 10^{-2} \left(2.6738\times 10^{11}+5.34761\times 10^{11} \lambda +2.6738\times 10^{11} \lambda ^2\right)}{(1 +\lambda )^2 \left(3.16388\times 10^{10}+1.27324\times 10^{10}\, \Sigma\right)}\times\nonumber\\&\times&\text{ProductLog}\left[\frac{3.048\times 10^{-20} N^2 \Delta ^{4\lambda } \left(3.16\times 10^{10}+1.27\times 10^{10}\,\Sigma\right)}{\alpha  \delta^2(1+\lambda )^2}\right],\label{nalpha2}
\end{eqnarray}
where we have defined new parameters as
\begin{eqnarray}
\Delta \equiv H/M,\,\delta=M/m_{\rm P},\,\Sigma\equiv \log\left[\frac{2.00 \times 10^{-5} \Delta ^{1+2\lambda }}{\delta }\right].
\end{eqnarray}
\begin{figure}[!h]	
	\includegraphics[width=8.3cm]{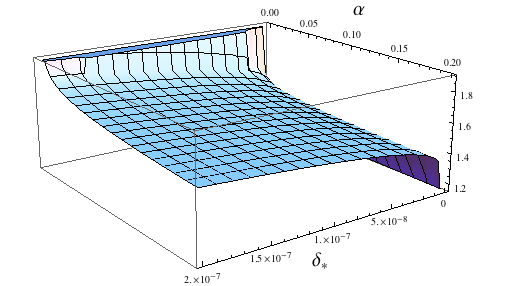}
	\includegraphics[width=8.3cm]{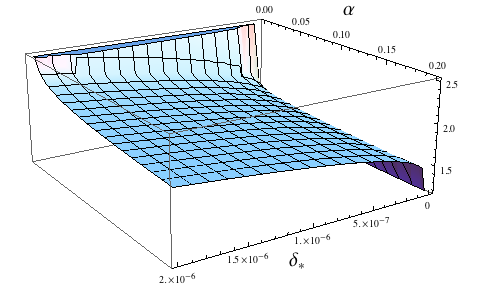}
	\centering
	\caption{Figures show the behavior of $n$ against $\delta_{\star}$ and $\alpha$ given by Eq.(\ref{nalpha2}). A vertical axis represents values of $n$. For the plots we have used $N=60$ (left panel) and $N=70$ (right panel).}
	\label{plothuc1}
\end{figure}
Having used an upper bound on the Hubble parameter during inflation observed by Planck, we can quantify how values of $n$ do depend on $\alpha$ and $\delta_{\star}$ during inflation. From Fig.\ref{plothuc1}. we find the behavior of $n$ against $\delta_{\star}$ and $\alpha$ where we have used $\Delta_{\star}=H_{\star}/M,\,\lambda=0.1$. Notice that 

%%%%%%%%%%%%%%%%%%%%%%%%%%%%%%%%%%
\subsection{Model II: $f(R)=R+\alpha R^{2} + \beta R^{2}\log(R/M^{2})$}
%%%%%%%%%%%%%%%%%%%%%%%%%%%%%%%%%
In this second model, we consider a function $f(R)$ of the form
\begin{eqnarray}
f(R)=R+\alpha R^{2} + \beta R^{2}\log(R/M^{2}).
\end{eqnarray}
We can estimate $F(R)=\partial f(R)/dR$ by assuming $H\gg M$ during inflation to obtain
\begin{eqnarray}
F(R)\approx 12 H^2 \left(\frac{H}{M}\right)^{2 \lambda } \left(2 \alpha +\beta +2 \beta \log\left[12 \left(\frac{H}{M}\right)^{2+2 \lambda }\right]\right).
\end{eqnarray}
Considering Eq.(\ref{a32}), then the power spectrum of curvature perturbation reads
\begin{eqnarray}
    \mathcal{P}_{\mathcal{R}} \approx  \frac{1}{36 \pi m_{\rm P}^2 \left(2 \alpha +\beta +2 \beta  \log\left[12 \left(\frac{H}{M}\right)^{2 (1+\lambda )}\right]\right)}\frac{1}{(1+\lambda)^{2}\epsilon^{2}_{1}}\,.\label{PRmod2}
\end{eqnarray}
In this model, therefore, $\epsilon_3$ reads
\begin{eqnarray}
    \epsilon_3=\frac{\dot{H} \left(4 \alpha  \lambda +2 \alpha +10 \beta  \lambda +7 \beta +2 (2 \beta  \lambda +\beta ) \log \left(12 H^{2 \lambda +2} M^{-2 (\lambda +1)}\right)\right)}{H^2 \left(2 \alpha +3 \beta +2 \beta  \log \left(12 H^{2 \lambda +2} M^{-2 (\lambda +1)}\right)\right)}+\frac{\ddot{H}}{H \dot{H}}\,.\label{ep3}
\end{eqnarray}
Assuming slow-roll approximations, the terms containing $\ddot{H}$ can be ignored and then the relation between $\epsilon_3$ and $\epsilon_1$ reads
\begin{eqnarray}
    \epsilon_3=\epsilon_1 \left(-1 -2 \lambda-\frac{4 \beta  (\lambda +1)}{2 \alpha +3 \beta +2 \beta  \log \left(12 M^{-2 \lambda -2} H(t)^{2 \lambda +2}\right)}\right)\,,\label{ep3mod1}
\end{eqnarray}
where we have used Eq.(\ref{ep1ep2}) and Eq.(\ref{ep3mod1}) together with Eq.(\ref{a31}). Hence, we have
\begin{eqnarray}
 n_{\mathcal{R}} - 1 &\simeq& -2(\lambda+2)\epsilon_1+2\epsilon_2-2\epsilon_3 \ ,\\
 n_{\mathcal{R}}&=&1-\frac{4 \epsilon _1 \left(2 \alpha -2 \beta  \lambda +\beta +2 \beta  \log \left(12 H^{2 \lambda +2} M^{-2 (\lambda +1)}\right)\right)}{2 \alpha +3 \beta +2 \beta  \log \left(12 H^{2 \lambda +2} M^{-2 (\lambda +1)}\right)}\ ,
\end{eqnarray}
where we have defined a new parameter $\Delta \equiv H/M$. For simplicity, let us suppose that during inflation the expansion is de
Sitter (exponential) with a constant Hubble parameter. In terms of the number of efoldings, $\mathcal{P}_{\mathcal{R}},\,n_\mathcal{R}$ and $r$ read
\begin{eqnarray}
    \mathcal{P}_{\mathcal{R}} &\approx& \frac{N^2}{9 m_{\rm P}^2\pi  (1+\lambda )^2}\Big(2 \alpha +\beta +2 \beta  \text{Log}\Big[12 \left(\Delta\right)^{2 (1+\lambda )}\Big]\Big)^{-1},\label{PR33m2}\\
    n_\mathcal{R}&\approx&1-\frac{2}{N}+\frac{4 \beta  (\lambda +1)}{N \left(2 \alpha +2 \beta  \log \left(12 \Delta ^{2 \lambda +2}\right)+3 \beta \right)}, \label{finalnsm2} \\
    r&\approx&\frac{12 (\lambda +1)^2}{N^2}. \label{finalrm2}
\end{eqnarray}
\begin{figure}[!h]	
	\includegraphics[width=8.2cm]{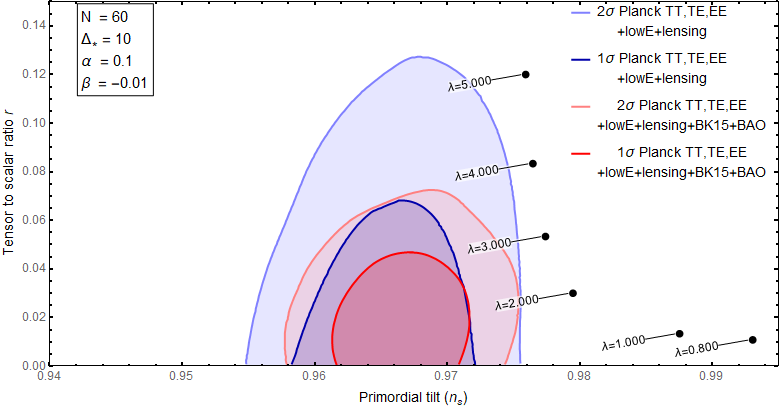}
	\includegraphics[width=8.2cm]{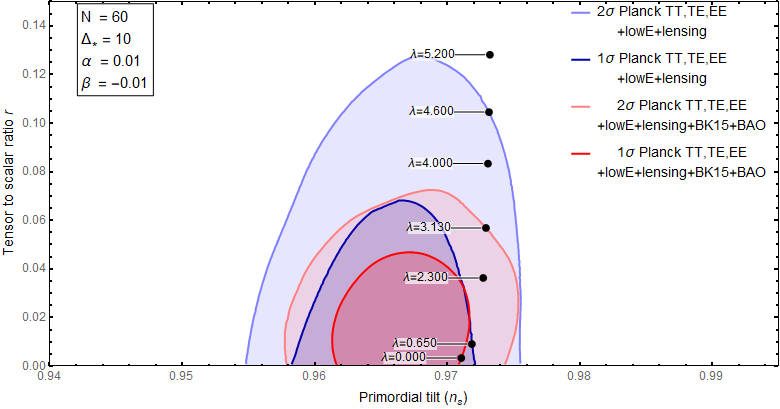}
	\includegraphics[width=8.2cm]{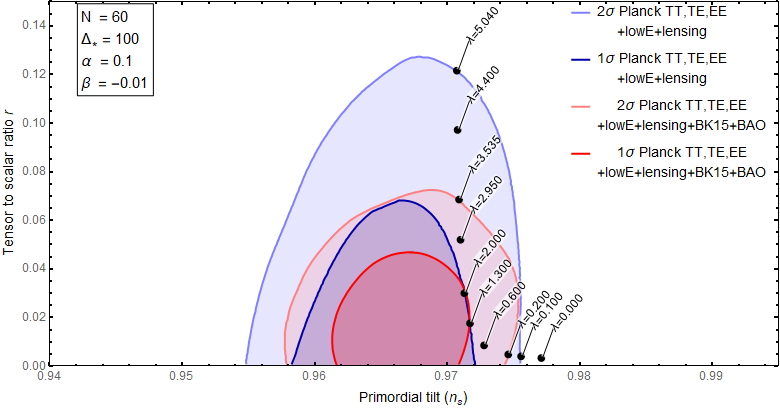}
	\includegraphics[width=8.2cm]{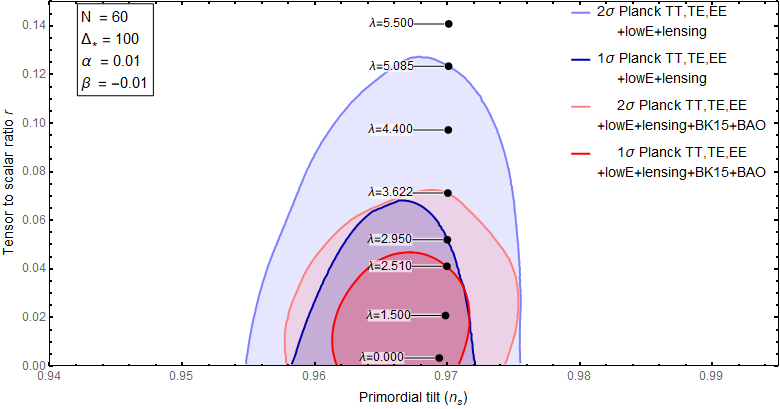}
	\centering
	\caption{We compare the theoretical predictions in the $(r-n_{s})$ plane for different values of $\lambda$ using $\Delta_{\star}=10,\,100,\,\alpha=0.1,\,0.01,\,\beta=-0.01,\,N=60$. (upper-left panel); $\Delta_{\star}=100,\,\alpha=0.01,\,\beta=-0.01,\,N=60$ (upper-right panel); $\Delta_{\star}=10,\,\alpha=0.1,\,\beta=-0.01,\,N=60$ (lower-left panel) and $\Delta_{\star}=100,\,\alpha=0.01,\,\beta=-0.01,\,N=60$ (lower-right panel) with Planck’15 results for TT, TE, EE, +lowE+lensing and +BK15+BAO.}
	\label{nsrmod2}
\end{figure}

We find that the above parameters reduce to those of the Starobinsky model when $\beta=0=\lambda$. We now compare our predicted results with Planck 2018 data. We find from Fig.(\ref{nsrmod2}) for $N=60$ that the predictions are inconsistent with the Planck’15 results for TT, TE, EE, +lowE+lensing and +BK15+BAO at two sigma confidence level for $\Delta_{\star}=10,\,\alpha=0.1,\,\beta=-0.01$. However, the predictions lie well inside the two-sigma regions for $\Delta_{\star}=10,\,\alpha=0.01,\,\beta=-0.01$. Using $\Delta_{\star}=100$, we discover that the results lie inside the two-sigma regions for $\alpha=0.1,\,\beta=-0.01$ and are in good agreement with TT, TE, EE, +lowE+lensing+BK15+BAO for $\alpha=0.01,\,\beta=-0.01$. We conclude that in order to have the predictions fit well inside the one-sigma regions of the Planck 2018 data we need either $\Delta_{\star}\gg {\cal O}(10)$ or $\,\alpha=|\beta|\ll 0.1$.

Additionally, for $N=60$, we discover that the predictions are in excellent agreement with consistent with the Planck’15 results for TT, TE, EE, +lowE+lensing and TT, TE, EE, +lowE+lensing and +BK15+BAO at one sigma confidence level for $\Delta_{\star}=100,\,0\leq \lambda\leq 2.950$ and $\Delta_{\star}=100,\,0\leq \lambda\leq 2.510$, respectively. Notice that when setting $\beta=0$, we clearly obtain the results of the Starobinsky model. Using Planck 2018 data for $\mathcal{P}_{\mathcal{R}}$, we can solve for $\Delta$ to obtain
\begin{eqnarray}
\Delta={\rm Exp}\Bigg[\frac{\Big(\beta  (-1.14-1.14\lambda )^3+\alpha  (-0.79-0.79\lambda )^3\Big)}{\beta  (1+\lambda )^4} + \frac{4.21\times 10^6 N^2 (1+\lambda )}{m_{\rm P}^2 \beta  (1+\lambda )^4}\Bigg],\label{Del2}
\end{eqnarray}
\begin{figure}[!h]	
	\includegraphics[width=8.5cm]{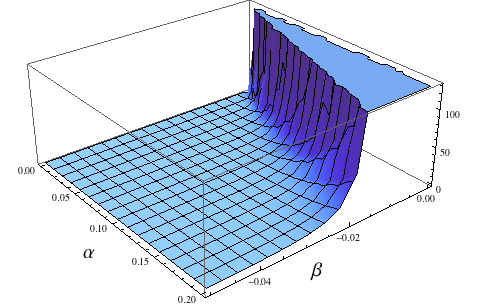}
	\includegraphics[width=8.1cm]{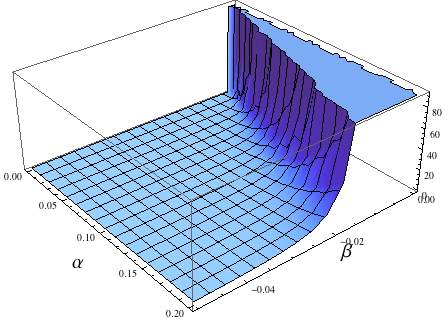}
	\centering
	\caption{Figures show the behavior of $\Delta$ against $\alpha$ and $\beta$ given by Eq.(\ref{Del2}). A vertical axis represents values of $\Delta$. For the plots we have used $N=60,\,\lambda=0.01$ (left panel) and $N=60,\,\lambda=0.1$ (right panel).}
	\label{plothu2}
\end{figure}
It is worth noting that when $\beta=0$ a term $\Delta$ disappears. We verify the behavior of $\Delta$ during inflation as illustrated in Fig.\ref{plothu2}. From figures, we find that in order to obtain sizeable values of $\Delta_{\star}$, e.g. $\Delta_{\star}\sim {\cal O}(10)-{\cal O}(100)$, we find that $\beta$ must be negative in a range $-1.0\ll \beta <-0.1$ together with $\alpha\ll 0.5$.
%%%%%%%%%%%%%%%%%%%%%%%%%%%%%%%%%%%%%%%%%%%%%
\subsection{Model III: Einstein-Hu-Sawicki}
%%%%%%%%%%%%%%%%%%%%%%%%%%%%%%%%%%%%%%%%%%%%
In the last model, we consider the Hu-Sawicki model in which the $f(R)$ function is of the form
\begin{eqnarray}
f(R)=R-M^{2}\frac{c_{1}\Big(R/M^{2}\Big)^{n}}{c_{2}\Big(R/M^{2}\Big)^{n}+1},
\end{eqnarray}
where we have fixed $n=2$ in this present examination. Then $F$ defined by $F(R)=\partial f(R)/\partial R$ can be written by
\begin{eqnarray}
   F(R)=1+\frac{2 c_1 c_2 R^3}{M^6 \left(\frac{c_2 R^2}{M^4}+1\right)^2}-\frac{2 c_1 R}{M^2 \left(\frac{c_2 R^2}{M^4}+1\right)}.
\end{eqnarray}
Therefore, $\epsilon_3$ reads
\begin{eqnarray}
    \epsilon_3=\frac{\dot{H}}{H^2}\left(\frac{12 (\lambda +1) M^4}{144 c_2 H^4 \left(\frac{H}{M}\right)^{4 \lambda }+M^4}-\frac{4 (\lambda +1) M^4}{M^4-432 c_2 H^4 \left(\frac{H}{M}\right)^{4 \lambda }}-6 \lambda -7\right)+\frac{\ddot{H}}{H \dot{H}}\,.\label{ep3}
\end{eqnarray}
Assuming slow-roll approximations, the terms containing $\ddot{H}$ can be ignored and then the relation between $\epsilon_3$ and $\epsilon_1$ reads
\begin{eqnarray}
    \epsilon_3=\epsilon_1 \left(-\frac{12 (\lambda +1)}{144 c_2 \left(\frac{H}{M}\right)^{4 \lambda +4}+1}+\frac{4 (\lambda +1)}{1-432 c_2 \left(\frac{H}{M}\right)^{4 \lambda +4}}+6 \lambda +7\right)\,,\label{ep3ep1}
\end{eqnarray}
where we have used Eq.(\ref{ep1ep2}) and Eq.(\ref{ep3}) together with Eq.(\ref{a31}). Hence, we have
\begin{eqnarray}
 n_{\mathcal{R}} - 1 &\simeq& -2(\lambda+2)\epsilon_1+2\epsilon_2-2\epsilon_3 \ ,\\
 n_{\mathcal{R}}&=&1-2 (\lambda +2) \epsilon_1+2 (-\lambda -1) \epsilon_1\nonumber\\&& +2 \epsilon_1 \left(\frac{12 (\lambda +1)}{144 c_2 \Delta^{4 \lambda +4}+1}-\frac{4 (\lambda +1)}{1-432 c_2 \Delta^{4 \lambda +4}}-6 \lambda -7\right)\ ,
\end{eqnarray}
where we have defined a new parameter $\Delta \equiv H/M$. For simplicity, let us suppose that during inflation the expansion is de
Sitter (exponential) with a constant Hubble parameter. Here we assume that $H\approx H_{\star}={\rm constant}$ and then define a new parameter $\Delta_{\star}\equiv (H_{\star}/M)$ which is plausible during inflation. We consider Eq.(\ref{a32}) and then the power spectrum of curvature perturbation reads
\begin{eqnarray}
    \mathcal{P}_{\mathcal{R}} \approx \frac{1}{Q_s}\left(\frac{H}{2\pi}\right)^2\left(\frac{H}{M}\right)^{2\lambda} = \frac{1}{3\pi F}\left(\frac{H}{m_{\rm P}}\right)^{2}\left(\frac{H}{M}\right)^{2\lambda}\frac{1}{(1+\lambda)^{2}\epsilon^{2}_{1}}\,,
\end{eqnarray}
when we have inserted $Q_s$ defined in Eq.(\ref{4.27}). We will see that $c_{2}\lll 1$ allowing to estimate $F(R)\approx- 2 c_1 R/M^2$ since $R\gg M^{2}$ during inflation. With this approximation, we find
\begin{eqnarray}
    \mathcal{P}_{\mathcal{R}} \approx \frac{1}{72\pi {\tilde c_{1}}}\left(\frac{M}{m_{\rm P}}\right)^2\frac{1}{(1+\lambda)^{2}\epsilon^{2}_{1}}\,,\label{PR3}
\end{eqnarray}
where ${\tilde c_{1}}=-c_{1}$. Furthermore, we can recall equation (\ref{a44}) and use equation (\ref{ep1ep2}) to find the tensor-to-scalar ratio $r$:
\begin{eqnarray}
     r = 48\epsilon_2^2 = 48(\lambda+1)^2\epsilon_1^2.
\end{eqnarray}
Now, we obtain the expressions of $n_\mathcal{R}$, $\mathcal{P}_\mathcal{R}$ and $r$. Consequently. we can rewrite $n_\mathcal{R}$ and $r$ in terms of e-folds ($N$) because it is convenient to constrain the predictions of our model. To do so, we can use the relation between $\epsilon_1$ and $N$ given in Eq.(\ref{epN3}) at the time for which a Hubble radius crossing. Therefore, $\mathcal{P}_{\mathcal{R}},\,n_\mathcal{R}$ and $r$ read 
\begin{eqnarray}
    \mathcal{P}_{\mathcal{R}} &\approx& \frac{1}{18\pi {\tilde c_{1}}}\left(\frac{M}{m_{\rm P}}\right)^2\frac{N^{2}}{(1+\lambda)^{2}},\label{PR33}\\
    n_\mathcal{R}&\approx&1-\frac{2\lambda+3}{N}+\frac{1}{N}\Bigg(\frac{12 (\lambda +1)}{144 c_2 \Delta^{4 \lambda +4}+1}-\frac{4 (\lambda +1)}{1-432 c_2 \Delta^{4 \lambda +4}}-6 \lambda -7\Bigg), \label{finalns} \\
    r&\approx&\frac{12 (\lambda +1)^2}{N^2}. \label{finalr}
\end{eqnarray}
\begin{figure}[!h]	
	\includegraphics[width=8.2cm]{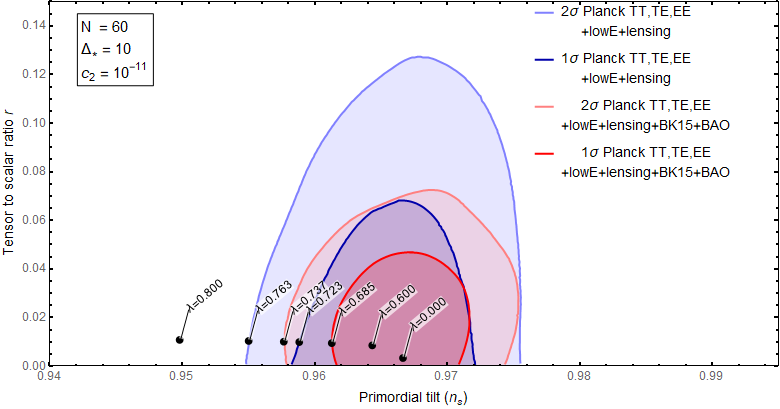}
	\includegraphics[width=8.2cm]{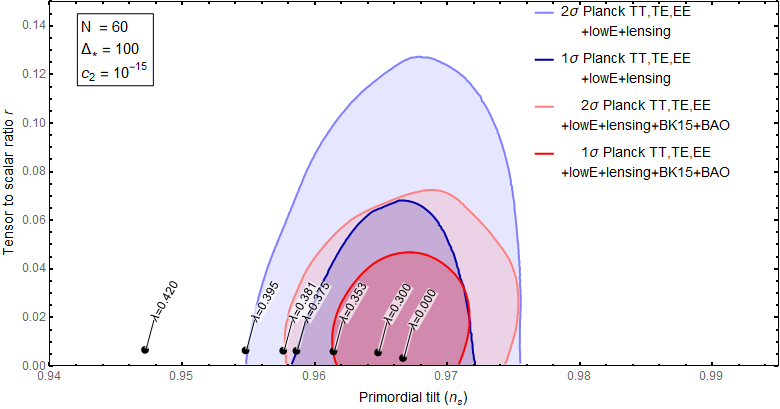}
	\includegraphics[width=8.2cm]{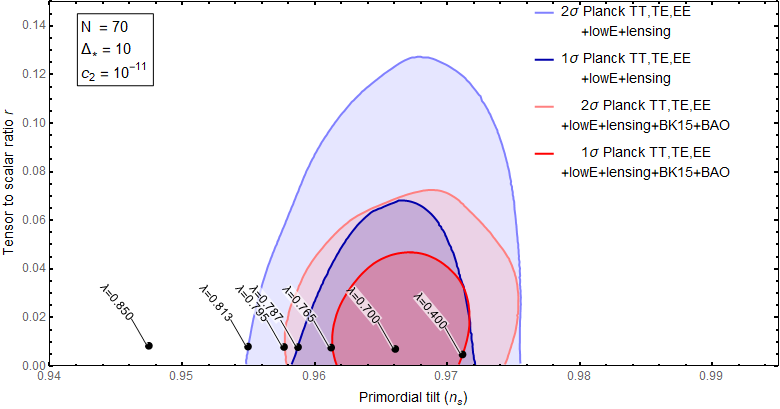}
	\includegraphics[width=8.2cm]{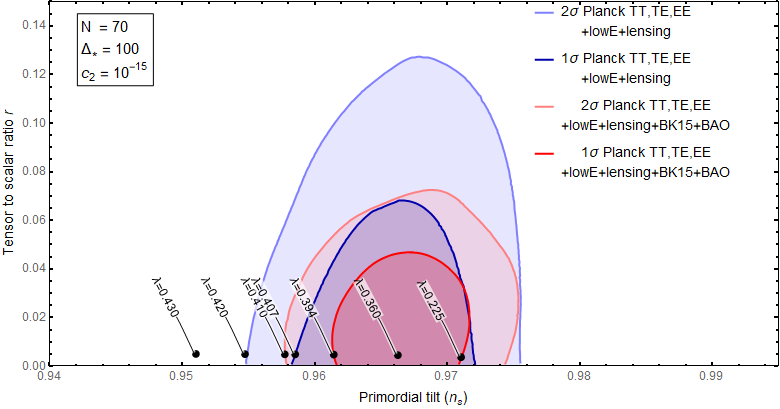}
	\centering
	\caption{We compare the theoretical predictions in the $(r-ns)$ plane for different values of $\lambda$ using $\Delta_{\star}=10, c_{2}=10^{-11}, N=60$ (upper-left panel); $\Delta_{\star}=100, c_{2}=10^{-15}, N=60$ (upper-right panel); $\Delta_{\star}=10, c_{2}=10^{-11}, N=70$ (lower-left panel) and $\Delta_{\star}=100, c_{2}=10^{-15}, N=70$ (lower-right panel) with Planck’15 results for TT, TE, EE, +lowE+lensing and +BK15+BAO.}
	\label{nsrmod3}
\end{figure}
We find that the above parameters reduce to those of the Starobinsky model when $c_{2}=0=\lambda$. We now compare our predicted results with Planck 2018 data. We find from
Fig.(\ref{nsrmod3}) for $N=60$ that the predictions are consistent with the Planck’15 results for TT, TE, EE, +lowE+lensing at two sigma confidence level for $\Delta_{\star}=10(100),\,c_{2}=10^{-11}(10^{-15})$ only when $\lambda\leq 0.763(0.395)$ with Planck’15 results for TT, TE, EE, +lowE+lensing and with the Planck’15 results for TT, TE, EE, +lowE+lensing+BK15+BAO at two sigma confidence level for $\Delta_{\star}=10(100),\,c_{2}=10^{-11}(10^{-15})$ only when $\lambda\leq 0.737(0.381)$. Additionally, for $N=70$, we discover that the predictions are consistent with the Planck’15 results for TT, TE, EE, +lowE+lensing at two sigma confidence level for $\Delta_{\star}=10(100),\,c_{2}=10^{-11}(10^{-15})$ only when $\lambda\leq 0.813(0.420)$ with Planck’15 results for TT, TE, EE, +lowE+lensing and with the Planck’15 results for TT, TE, EE, +lowE+lensing+BK15+BAO at two sigma confidence level for $\Delta_{\star}=10(100),\,c_{2}=10^{-11}(10^{-15})$ only when $\lambda\leq 0.795(0.410)$.

Interestingly, using an upper bound on the Hubble parameter during inflation reported by Planck 2018 \cite{Akrami:2018odb} allows us to determine $\delta_{\star}=M/m_{\rm P}$:
\begin{eqnarray}
    \frac{H_{\star}}{m_{\rm P}}=\Delta_{\star}\delta_{\star} <2.5\times 10^{-5}\,(95\% \,C.L.)\,\,\rightarrow\,\,\delta_{\star} <\frac{2.5}{\Delta_{\star}}\times 10^{-5}.\label{HUcon}
\end{eqnarray}
Using Planck 2018 data for $\mathcal{P}_{\mathcal{R}}$, we discover
\begin{eqnarray}
    {\tilde c_{1}} \approx \frac{3.53\times 10^{22}\,N^{2}\delta^{2}_{\star}}{4.20\times 10^{15}+8.40\times 10^{15}\lambda+4.20\times 10^{15}\lambda^{2}} .\label{c1t}
\end{eqnarray}
\begin{figure}[!h]	
	\includegraphics[width=8.2cm]{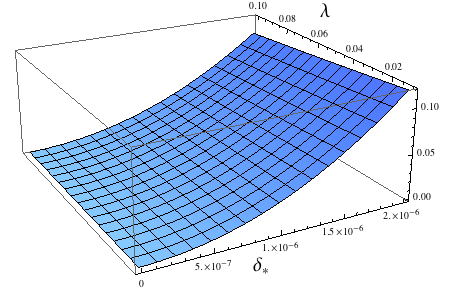}
	\includegraphics[width=7.5cm]{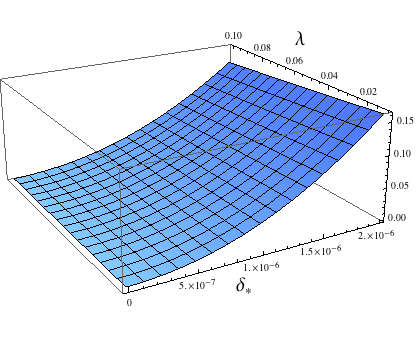}
	\centering
	\caption{Figures show the behavior of ${\tilde c_{1}}$ against $\delta_{\star}$ and $\lambda$ given by Eq.(\ref{c1t}). A vertical axis represents values of ${\tilde c_{1}}$. For the plots we have used $N=60$ (left panel) and $N=70$ (right panel).}
	\label{plothuc3}
\end{figure}
In order to figure out the behavior of ${\tilde c_{1}}$, we choose particular values of $N$ and $\lambda$ and make plots displayed in Fig.\ref{plothuc3}. We discover that in order to have parameter satisfying the Planck 2018 data, values of ${\tilde c_{1}}$ is (much) less than unity for $\Delta_{\star}=10\,(\Delta_{\star}=100)$. Interestingly, we can recover the predictions of the Starobinsky model, i.e. ${\tilde c_{1}}=1/6\approx 0.167$, when $\lambda\lll 1$.

%%%%%%%%%%%%%%%%%%%%
\section{Concluding remarks}
%%%%%%%%%%%%%%%%%%%%

In conclusion, we studied the $f(R)$ models of inflation in the context of gravity's rainbow theory. We have chosen three types of $f(R)$ models: $f(R)=R+\alpha (R/M)^{n},\,f(R)=R+\alpha R^{2}+\beta R^{2}\log(R/M^{2})$ and the Einstein-Hu-Sawicki model with $n,\,\alpha,\,\beta$ being arbitrary real constants. Here $R$ and $M$ being the Ricci scalar and mass scale, respectively. For all models, the rainbow function is written in the power-law form of the Hubble parameter. We presented a detailed derivation of the spectral index of curvature perturbation and the tensor-to-scalar ratio and compared the predictions of our results with latest Planck 2018 data. With the sizeable number of e-foldings and proper choices of parameters, we discovered that the predictions of all $f(R)$ models present in this work are in excellent agreement with the Planck analysis.

For $f(R)=R+\alpha (R/M)^{n}$ model, we discovered that values of $n$ cannot be any arbitrary. Using $N=60$, we observed that $n=2.235$ in order to have the results in agreement with the Planck 2018 data at the two sigma confident level. With $f(R)=R+\alpha R^{2} + \beta R^{2}\log(R/M^{2})$, we conclude for $N=60$ that in order to have the predictions fit well inside the one-sigma regions of the Planck 2018 data we need either $\Delta_{\star}\gg {\cal O}(10)$ or $\,\alpha=|\beta|\ll 0.1$ and found that $\beta$ has to be negative. Having considered the last model, the Einstein-Hu-Sawicky gravity, we observed that with the sizeable number of e-foldings and proper choices of parameters our predictions are in good agreement with the Planck’15 results for TT, TE, EE, +lowE+lensing and +BK15+BAO. Last but not the least, regarding our present work, the effects of rainbow functions on the structure of compact objects are also worth investigating, see e.g. \cite{EslamPanah:2017ugi,Hendi:2015vta}.

\section*{Acknowledgments}
The work of AW is financially supported by the Development and Promotion of Science and Technology Talents Project (DPST), the Institute for the Promotion of Teaching Science and Technology (IPST).

\appendix

\section{Scalar (Ricci) curvature in gravity' rainbow}
\label{apa} 
It is rather straight forward to derive the Ricci scalar of the theory. Let us first consider the Ricci curvature $R_{\mu\nu}$ and compute its $R_{00}$-component to obtain 
\begin{eqnarray}
R_{00}=-3 \left(\frac{\ddot{a}}{a}-\frac{\ddot{\tilde{g}}}{\tilde{g}}+\frac{\dot{a} \dot{\tilde{f}}}{a \tilde{f}}-\frac{\dot{\dot{\tilde{g}}\tilde{f}} }{ \tilde{g}\tilde{f}}-2\frac{ \dot{a} \dot{\tilde{g}}}{a \tilde{g}}+2 \left(\frac{\dot{\tilde{g}}}{\tilde{g}}\right)^2\right)\,,\label{R00}
\end{eqnarray}
while the $R_{ii}$-component reads 
\begin{eqnarray}
R_{ii}=\frac{a \dot{a} \tilde{f}\dot{\tilde{f}} +2 \dot{a}^2 \tilde{f}^2+a \ddot{a}\tilde{f}^2 }{\tilde{g}^2}-\frac{a^2 \tilde{f}\dot{\tilde{f}}  \dot{\tilde{g}}+6 a\dot{a} \tilde{f}^2 \dot{\tilde{g}}+a^2 \tilde{f}^2 \ddot{\tilde{g}}}{\tilde{g}^3}+\frac{4 a^2 \tilde{f}^2 \dot{\tilde{g}}^2}{\tilde{g}^4}\,.\label{Rii}
\end{eqnarray}
Clearly, when setting $\tilde{f}=1$ and $\tilde{g}=1$, the above results convert to those of the standard flat FLRW spacetime in the $f(R)$ theory. The Ricci scalar $R$ con be simply extracted by contracting $R_{\mu\nu}$ with $g_{\mu\nu}$ to obtain 
Ricci scalar
\begin{eqnarray}
R=g^{\mu\nu}R_{\mu\nu}=6 \tilde{f}^2 \left(\frac{\ddot{a}}{a}+\frac{\dot{a} \dot{\tilde{f}}}{a \tilde{f}}+\frac{\dot{a}^2}{a^2}\right)-6 \tilde{f}^2 \left(\frac{\ddot{\tilde{g}}}{\tilde{g}}+\frac{\dot{\tilde{f}} \dot{\tilde{g}}}{\tilde{f} \tilde{g}}+4\frac{\dot{a} \dot{\tilde{g}}}{a \tilde{g}}-3\frac{ \dot{\tilde{g}}^2}{\tilde{g}^2}\right).
\end{eqnarray}

%%%%%%%%%%%%%%%%%%%%%%%%%%%%%%%%%%%%%%%%
%%%%%%%%%%%%%%%%%%%%%%%%%%%%%%%%%%%%%%%%
\end{document}